\documentclass[10pt,twocolumn,superscriptaddress,aps,prx,longbibliography,showpacs,floatfix]{revtex4-2}
\usepackage{dcolumn}                 
\usepackage{times}
\usepackage{nicefrac}
\usepackage{amsmath}
\usepackage{amsfonts}
\usepackage{amssymb}
\usepackage{amsthm}
\usepackage{xcolor}
\usepackage{graphicx}
\newcommand*{\centt}[1]{\multicolumn{1}{c}{#1}}
\newcommand*{\cent}[1]{\multicolumn{1}{c}{$#1$}}
\newcolumntype{w}[1]{D{.}{.}{#1}}
\newcolumntype{x}{>{$}r<{$}}

\newcommand{\icm}{\mathrm{cm}^{-1}}

\newcommand{\etal}{\textit{et al.}\,}

\usepackage{hyperref}
\usepackage{MnSymbol}   
    
\definecolor{green}{rgb}{.0,.5,.0}

\definecolor{red}{rgb}{1.,.0,.0}

\definecolor{grey}{rgb}{0.5,0.5,0.5}

\allowdisplaybreaks

\def\sep{ }

\begin{document}

\preprint{Version 2.2}

\title{Fine and hyperfine splitting of the low-lying states of $^9$Be}

\author{Mariusz Puchalski}
\affiliation{Faculty of Chemistry, Adam Mickiewicz University, Uniwersytetu Pozna{\'n}skiego 8, 61-614 Pozna{\'n}, Poland}

\author{Jacek Komasa}
\affiliation{Faculty of Chemistry, Adam Mickiewicz University, Uniwersytetu Pozna{\'n}skiego 8, 61-614 Pozna{\'n}, Poland}

\author{Krzysztof Pachucki}
\affiliation{Faculty of Physics, University of Warsaw, Pasteura 5, 02-093 Warsaw, Poland}

\date{\today}

\begin{abstract}
We perform accurate calculations of energy levels as well as fine and hyperfine splittings of the lowest 
$^{1,3}\!P_J$, $^{3}\!S_1$, $^{3}\!P^e_J$, and $^{1,3}\!D_J$ excited states of the $^9$Be atom using
explicitly correlated Gaussian functions and report on the breakdown of the standard hyperfine 
structure theory. Because of the strong hyperfine mixing, which prevents the use of common hyperfine 
constants, we formulate a description of the fine and hyperfine structure that is valid for 
an arbitrary coupling strength and may have wide applications in many other atomic systems.  
\end{abstract}

\pacs{31.15.ac, 31.30.J-}
\maketitle


\section{Introduction}

The main drawback of atomic structure methods based on the nonrelativistic wave function 
represented as a linear combination of determinants of spin-orbitals
(Hartree-Fock, configuration interaction, multi-configurational self-consistent field, etc.) 
is the difficulty in providing results with reliably estimated uncertainties. 
While the accuracy of nonrelativistic energy can be assessed by increasing the space of electronic configurations, 
the irregular convergence of matrix elements, especially involving singular operators, such as those for relativistic or quantum electrodynamics (QED) corrections, 
often does not allow for presenting any uncertainties. Therefore, this deficiency limits the use of these methods in the high-accuracy-demanding applications, 
e.g. testing quantum electrodynamics \cite{Nortershauser:15,Pachucki:17,Alighanbari:20}, 
determination of the nuclear charge radii \cite{Sanchez:06,Sturm:14,Manovitz:19,Maass:19}, 
the nuclear electromagnetic moments  \cite{Stone:15,Puchalski:21}, or physical constants \cite{CODATA:18}, 
and searching for new physics \cite{Safronova:18}.

On the other hand, an alternative approach based on the Dirac-Coulomb Hamiltonian,
with the wave function represented as a determinant of four-component spin-orbitals of positive energy, 
can reach reasonably convergence on relativistic energies and matrix elements \cite{Derevianko0}. But so far there is no
formulation of QED theory on the top of Dirac-Coulomb Hamiltonian with projection on positive one-electron energies.
Therefore numerical convergence does not say much about uncertainties due to omitted QED effects,
including those related to negative energy orbitals. Another problems arise when the hyperfine effects are not negligible compared to the fine structure splitting.
In the previous works on this topic (e.g. \cite{Derevianko1, Derevianko2}), it has been demonstrated the hyperfine mixing of fine structure levels can be satisfactorily accounted for by the second-order perturbation theory. However, when the hyperfine splitting is of the same order or even larger than the fine structure splitting, the perturbative approach fails, and one can no longer use the standard $A_J$ and $B_J$ hyperfine parameters.

It is desirable therefore to develop tools which provide  high and controlled accuracy, 
like those based on nonrelativistic QED (NRQED) theory and representation of the nonrelativistic wave function in terms of explicitly correlated basis functions, 
e.g. exponential, Hylleraas, or Gaussian (ECG) ones. The controlled accuracy is achieved by means 
of a full variational optimization of the wave function and by transformation of singular operators 
to an equivalent but more regular form. The price paid for using the explicitly correlated functions is the rapid increase 
in the complexity of calculations with each additional electron; therefore, application of these functions  
has so far been limited to few-electron systems only. 

Before passing to the main topic, which is the 4-electron beryllium (Be) atom, let us briefly describe recent advances 
in the calculation of 1-, 2-, and 3-electron atomic systems. Hydrogenic systems are the only ones in which theoretical
predictions including QED effects are sufficiently accurate to determine the nuclear (proton, deuteron)
charge radius from the measured transition frequencies \cite{CODATA:18}. Being apparently simple,
hydrogenic systems are a cornerstone for the implementation of QED in bound states, which relies on 
the expansion of binding energy in powers of the fine structure constant $\alpha\sim 1/137$. 
Similarly, for two- and more-electron systems, one also performs expansion in $\alpha$, 
as long as the nuclear charge $Z$ is not too large. This allows description of an atomic system in terms of the successively smaller effects,
i.e. nonrelativistic energy $\alpha^2$, relativistic correction $\alpha^4$, leading QED of order $\alpha^5$, and so on. 
For the helium atom all these expansion terms are calculated up to the order $\alpha^6$ \cite{Yerokhin:10}, 
with some states up to the order $\alpha^7$ \cite{Patkos:21}. Such high order calculations are 
feasible with explicitly correlated exponential basis functions, for which analytic integrals 
are known. 
Atomic systems with three electrons present greater difficulty for the accurate calculation 
of their energy levels despite obtaining very precise wave functions with explicitly correlated 
Hylleraas or ECG functions. 
Nevertheless, several highly accurate results have been obtained for Li and Be$^+$, including 
isotope shifts for the charge radii determination \cite{Sanchez:06,Northershauser:09,Krieger:12}, 
and fine \cite{Puchalski:14} and hyperfine \cite{Puchalski:13b} splitting. The fine structure 
splitting of the lithium $2^2\!P_J$ state with the inclusion of $O(\alpha^6)$ QED corrections
\cite{Puchalski:14,Wang:17} agrees well with even more accurate experimental values 
\cite{Brown:13,Li:20}, while current theoretical predictions for the $^{6/7}$Li ground state 
hyperfine splitting are limited by insufficient knowledge of the nuclear structure, and
not by the atomic structure theory. 

The experience gained from the above-mentioned systems can be exploited to a large extent 
in four-electron systems, but these ones are much more demanding in calculations. 
The attempts employing Hylleraas wave functions \cite{Busse:98,King:11,Sims:11} are limited to 
nonrelativistic energy because of the lack of effective methods of evaluation of "relativistic" four-electron integrals. 
Therefore, our choice is the use of the ECG method, which performs very well for both nonrelativistic and relativistic contributions \cite{Komasa:95,Busse:98,Komasa:01a,Komasa:01b,Komasa:02a,Komasa:02b,Pachucki:04,PK06b,Stanke:07a,Stanke:07b,Stanke:09,Chen:09,Bunge:10,King:11,Sims:11,Sharkey:11,Chen:12,Bubin:12,Puchalski:13a,Puchalski:14a,Sharkey:14,Stanke:19,Kedziorski:20}. 
Nonetheless, we note that because Gaussian-type wave functions do not satisfy the Kato cusp condition,
the complete calculation of the $\alpha^6$ correction is still unfeasible. This unsolved problem limits 
the current capabilities of the ECG method in the application to three- and more-electron systems.
After all, the ECG functions are so far the best suited for the four-electron systems 
and such an application to the Be atom will be presented here.

In our previous works on Be we have obtained accurate energies for the ground $2s^2\,^{1}\!S_0$ 
and the excited $2s2p\,^{1}\!P_1$ states \cite{Puchalski:13a}, the difference of which at $42\,565.441(11)\,\icm$, 
agrees well with the experimental value of $42\,565.450\,1(13)\,\icm$ obtained by Cook {\em et al.} \cite{Cook:18} and with later 
calculations \cite{Hornyak:19}. Last year, two more transitions were measured to a high accuracy---the wavenumber of the 
$2s^2\,^1\!S_0 - 2s3d\,^1\!D_2$ line, equal to $64\,428.403\,21(55)\,\icm$, and of the 
$2s2p\,^1\!P_1 - 2s3d\,^1\!D_2$ line, equal to $21\,862.952\,9(14)\,\icm$, were reported 
by Cook {\em et al.} \cite{Cook:20}. In this case, no theoretical results at an adequate 
level of accuracy have been calculated yet. Moreover, 
in the beryllium atom, of particular theoretical interest is the lowest $2s2p\,^3\!P$ excited state,
because it is metastable. So far though, its energy has not been measured and calculated to 
such a high accuracy as for  the $2s2p\,^1\!P$ state. An old but the most accurate experimental excitation energy from the ground 
to the $2s2p\,^3\!P_1$ level equal to $21\,978.92(1)~\icm$ \cite{Bozman:53} is in agreement 
with the less accurate recent theoretical value of $21\,978.2(11)~\icm$ by Kedziorski {\em et al.}
\cite{Kedziorski:20}. Quite recently, the hyperfine splitting of the $2s2p\,^{3}\!P$ state 
has been accurately calculated and, with the help of the 50-years old measurements by Blachman \cite{Blachman:67},
has been employed to determine the most accurate value of the electric quadrupole moment of $^9$Be
\cite{Puchalski:21}, but it is  in disagreement with all previous determinations.
Moreover, all the other Be energy levels lying below the ionization threshold of $75\,192.64\,\icm$ have large
uncertainties, being in the range $0.01-0.2\,\icm$ \cite{Kramida:20}, and there are no corresponding
accurate theoretical results to compare with.

The purpose of the present work is therefore to significantly advance the theoretical description of the lowest excited 
states with different internal symmetries. 
Namely, we focus on states with non-vanishing spin or orbital angular momentum
and verify the previous literature results, which were obtained using either the ECG method
or methods based on one-electron approximation. More precisely,
we report on ECG calculations for the six lowest excited states of the $^9$Be atom: 
$^{1,3}\!P_J$, $^{3}\!S_1$, $^{3}\!P^e_J$, and $^{1,3}\!D_J$, including their fine and hyperfine
splittings. Due to a significant hyperfine mixing, the standard hyperfine structure formulation 
in terms of $A_J$ and $B_J$ coefficients is not adequate in some cases. For this reason we have 
introduced a combined fine--hyperfine structure formalism that naturally accounts for 
an arbitrary mixing between fine and hyperfine levels. Moreover,
in order to unify the description of atomic wave functions of different symmetries,
we have introduced in this work a Cartesian angular momentum representation, 
which is tailored for use with many-electron explicitly correlated basis functions
and which simplifies evaluation of matrix elements.

\begin{figure}[!htb]
\includegraphics[scale=0.65]{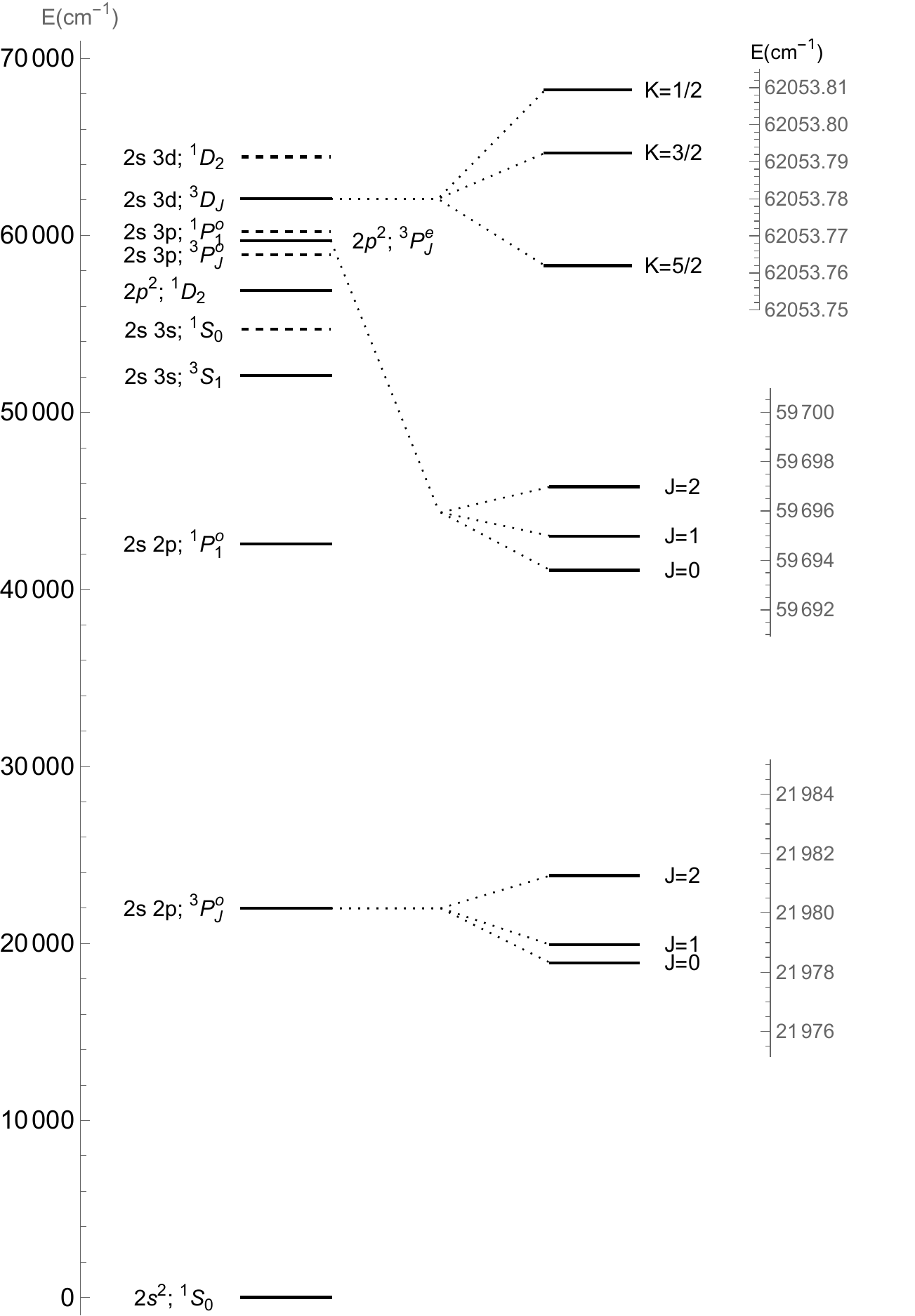}
\caption{The lowest energy levels of theBe atom. 
The levels for which the energy is evaluated in this work are drawn in solid lines and
the remaining ones in dashed line. The fine structure of the triplet states of our interest
is also shown, $\vec J = \vec L+\vec S, \vec K = \vec I+\vec S$.}
\label{Flevles}
\end{figure}

\section{Theoretical framework}

In the  calculations  of the energy levels of  few-electron light atomic systems with a well controlled accuracy,
we employ the expansion in the fine-structure constant $\alpha \approx 1/137$
\begin{equation}
E(\alpha) = E^{(2)}+E^{(4)} + E^{(5)} + E^{(6)} + \ldots,
\label{eq:expansion}
\end{equation}
where $E^{(n)}\sim m\, \alpha^n$
and some expansion terms may include finite powers of $\ln\alpha$. Uncalculated higher order terms 
will be estimated from the corresponding expansion terms in the hydrogenic limit, while the numerical accuracy
is controlled by the varying the number of terms in the highly optimized explicitly correlated wave  function.
\subsection{Nonrelativistic energy}
The leading term $E_0 \equiv E^{(2)}$ is obtained from the non-relativistic Hamiltonian in the center-of-mass system ($\vec p_N = -\sum_a \vec p_a$) 
by solving the Schr{\" o}dinger equation (in natural units)
\begin{align}
H^{(2)} \Psi &= E_0  \Psi \\
H^{(2)} &= \frac{\vec p^2_N}{2\,m_N} + \sum_a \frac{\vec p^2_a}{2\,m} - \sum_a \frac{Z\,\alpha}{r_a} + \sum_{a<b} \frac{\alpha}{r_{ab}}
\label{H2}
\end{align} 
where $Z$ is the nuclear charge, and
$m$ and $m_N$ are the electron and nuclear masses, respectively. In this work, the effects of the finite nuclear mass (recoil) are included in $H^{(2)}$. This is in contrast to the perturbative approach based on additional expansion  in the electron-nucleus mass ratio, which is particularly useful in the isotope shift calculations \cite{Puchalski:14a}. Once the wave function $\Psi$ is determined, all the corrections to the nonrelativistic energy $E_0$ can be expressed in terms of expectation values $\langle\Psi|\ldots|\Psi\rangle \equiv \langle \ldots \rangle$ of known operators.  

\subsection{Leading order corrections}
The leading relativistic $E^{(4)}$ correction is calculated as the mean value of the Breit-Pauli Hamiltonian \cite{Bethe:77}.
For convenience, we split this Hamiltonian, according to its inner composition, into three parts: 
the no-spin (ns), the fine-structure (fs), and the hyperfine-structure (hfs) part
\begin{equation}
\label{H4}
H^{(4)} = H_{\rm ns}^{(4)} + H_{\rm fs}^{(4)} + H_{\rm hfs}^{(4)}\,. 
\end{equation}
The spin-independent part in its explicit form is 
\begin{align} \label{H4ns}
H^{(4)}_{\rm ns} =&\ \sum_a \Big[ -\frac{\vec p^4_a}{8\,m^3} + \frac{\pi\,Z\,\alpha}{2\,m^2}\,\delta^3(r_{a}) \\
&\qquad + \frac{Z\,\alpha}{2\,m\,m_N}\, p_a^i\, \biggl(\frac{\delta^{ij}}{r_{a}}+\frac{r^i_{a}\,r^j_{a}}{r^3_{a}}
\biggr)\, p_N^j  \Big] \nonumber \\
& + \sum_{a<b} \Bigg[\frac{\pi\,\alpha}{m^2}\, \delta^3(r_{ab}) - \frac{\alpha}{2\,m^2}\, p_a^i\,\biggl(\frac{\delta^{ij}}{r_{ab}}+\frac{r^i_{ab}\,r^j_{ab}}{r^3_{ab}}
\biggr)\, p_b^j  \Bigg] \nonumber  .
\end{align}

The part related to the fine-structure effects, containing the vector $\vec{\sigma}_a$ 
of Pauli spin matrices of electron $a$, can be expressed as follows
\begin{align}
H^{(4)}_{\rm fs} =&\ \sum_a\,\frac{Z\,\alpha}{4}\,\vec{\sigma}_a \cdot \Big[ \frac{(g-1)}{m^2} 
\frac{\vec{r}_a}{r_a^3}\times\vec{p}_a  - \frac{g}{m\,m_{\rm N}}\,\frac{\vec r_a}{r_a^3}\times\vec p_{\rm N} \Big]  \nonumber \\
& +\sum_{a\neq b}\, \frac{\alpha}{4\,m^2}\,
\vec{\sigma}_a \cdot \Big[g\,\frac{\vec{r}_{ab}}{r_{ab}^3} \times\vec{p}_b-(g-1)\,\frac{\vec{r}_{ab}}{r_{ab}^3} \times\vec p_a\Big] \nonumber \\
& - \sum_{a<b} \frac{3\,\alpha\,g^2}{16\,m^2}\,\sigma_a^i\,\sigma_b^j\,\biggl(
\frac{r_{ab}^i\,r_{ab}^j}{r_{ab}^5} - \frac{\delta^{ij}}{3\,r_{ab}^3}\biggr)\,, \label{EHfs}
\end{align}
where $g$ is the free electron $g$-factor, which accounts for one-loop QED corrections.
Finally, the leading order Hamiltonian for the hyperfine splitting,
containing the nuclear spin $\vec{I}$, reads
\begin{align}\label{EHhfs}
H^{(4)}_{\rm hfs} = \sum_a &
\biggl[\frac{1}{3}\,\frac{Z\,\alpha\,g\,g_{\rm N}}{m\,m_{\rm N}}\,
\vec \sigma_a\cdot\vec I\,\pi\,\delta^3(r_a)\nonumber \\ 
&+\frac{Z\,\alpha\,g_\mathrm{N}}{2\,m\,m_\mathrm{N}}\,\vec I\cdot\frac{\vec r_a}{r_a^3}\times\vec p_a
 -\frac{Z\,\alpha\,(g_\mathrm{N}-1)}{2\,m_\mathrm{N}^2}\,\vec I\cdot\frac{\vec r_a}{r_a^3}\times\vec p_\mathrm{N}\nonumber \\ 
&+\frac{3\,Z\,\alpha\,g\,g_\mathrm{N}}{8\,m\,m_\mathrm{N}}\,\sigma_a^i\,I^j\,
\biggl(\frac{r_a^i\,r_a^j}{r_a^5} - \frac{\delta^{ij}}{3\,r_a^3}\biggr)\nonumber \\ 
&-\frac{\alpha\,Q_\mathrm{N}}{2}\,\biggl(\frac{r_a^i\,r_a^j}{r_a^5} 
- \frac{\delta^{ij}}{3\,r_a^3}\biggr)\,\frac{3\,I^i\,I^j}{I\,(2\,I-1)}\biggr]\,.
\end{align}
Here, $Q_\mathrm{N}$ is the electric quadrupole moment of the nucleus, 
and $g_{\rm N}$ is the nuclear $g$-factor defined by
\begin{equation}
g_{\rm N} = \frac{m_{\rm N}}{Z\,m_{\rm p}}\,\frac{\mu}{\mu_{\rm N}}\,\frac{1}{I}\,.
\end{equation}
%
\subsection{Higher order corrections}

\subsubsection{Centroid energy}
Higher-order corrections, $E^{(n)}$ with $n>4$, are usually much smaller than $E^{(4)}$
because they contain higher powers of $\alpha$. The explicit form of the $m \alpha^5$ terms is given by
\begin{eqnarray}\label{eq:HQED}
E^{(5)} &=& \frac{4\,Z\,\alpha^2}{3\,m^2}\,\left[\frac{19}{30}+\ln(\alpha^{-2}) - \ln k_0\right]\,
\sum_a\,\langle \delta^3(r_a) \rangle  \nonumber \\ 
&& + \frac{\alpha^2}{m^2} \left[\frac{164}{15}+\frac{14}{3}\,\ln\alpha
\right]\,\sum_{a<b} \,\langle \delta^3(r_{ab}) \rangle  \\
&& -\frac{7}{6\,\pi} \,m\,\alpha^5 \sum_{a<b} \,\biggl \langle P\left(\frac{1}{m\,\alpha\,r_{ab}^3} \right) \biggr \rangle\,, 
\end{eqnarray}
where $\ln k_0$ is the Bethe logarithm \cite{Bethe:47,Bethe:77}, and $P(1/r^3)$ is 
the Araki-Sucher term \cite{Araki:57,Sucher:58}.

A complete set of operators for the quantum electrodynamic $m \alpha^6$ correction to energy levels of light atoms 
has been derived recently \cite{Patkos:19}. However, due to the lack of a computational method suitable for 
a four-electron wave function, we use the following approximate formula which includes only the leading term
related to the hydrogenic Lamb shift
\begin{equation}
E^{(6)} \approx \frac{\pi\,Z^2\,\alpha^3}{m^2}\,\left[\frac{427}{96} - 2 \ln(2) \right] \sum_a \langle \delta^3(r_a) \rangle
\end{equation}
and estimate its uncertainty to be  about 25\%. This estimation is based on the former calculation 
of $m \alpha^6$ correction to helium energy levels \cite{bla}. 

\subsubsection{Fermi contact interaction}
\label{sec:epsilon}
The hyperfine Hamiltonian in Eq.~(\ref{EHhfs}) represents the leading hyperfine interactions, 
but there are also other small corrections which contain higher powers of the fine structure constant $\alpha$.
Because most of them are proportional to the Fermi contact interaction, we account for them
in terms of the $1+\epsilon$ factor multiplying the first term of the $H^{(4)}_{\rm fs}$ Hamiltonian
\begin{align}
(1+\epsilon)\, \frac{1}{3}\,\frac{Z\,\alpha\,g\,g_{\rm N}}{m\,m_{\rm N}}\,\sum_a \vec \sigma_a\cdot\vec I\,\pi\,\delta^3(r_a)\,.
\end{align}
Below, we briefly describe contributions included in the $\epsilon$ factor.

The $\mathcal{O}(\alpha)$ correction is analogous to that in hydrogenic systems \cite{Eides:01} 
and is due to the finite nuclear size and the nuclear polarizability. It is given by 
\cite{Eides:01, Puchalski:09}
\begin{align}
H^{(5)}_Z =&\  \bigl[-2\,Z\,\alpha\,m\,r_Z\bigr]\,
\vec I\cdot \sum_a \frac{2}{3}\,\frac{Z\,\alpha\,g_{\rm N}}{m\,m_{\rm N}}\,\vec \sigma_a\,\pi\,\delta^3(r_a)\,,
\end{align}
where $r_Z$ is a kind of effective nuclear radius called the Zemach radius. 
Disregarding the inelastic effects, this radius can be written in terms of 
the electric charge $\rho_E$ and magnetic-moment $\rho_M$ densities as
\begin{align}
r_Z = \int d^3 r\,d^3r'\,\rho_E(r)\,\rho_M(r')\,|\vec r-\vec r'|.
\end{align}
Nevertheless, the inelastic, i.e. polarizability, corrections can be significant, but 
because they are very difficult to calculate, they are usually neglected. 
In this work we account for possible inelastic effects by employing $r_Z = 4.07(5)\;{\rm fm}$
achieved from a comparison of very accurate calculations of hfs in $^9$Be$^+$ with 
the experimental value~\cite{Puchalski:09}. Because this correction is also proportional 
to the contact Fermi interaction, we represent it in terms of $\epsilon_{\rm Z} = -0.000\,615$.
There is a small recoil correction at the same order of $\alpha$, for which we refer to \cite{Puchalski:17, Puchalski:21} and it contributes $\epsilon_{\rm rec} = -0.000\,011$.

Next, there are radiative and relativistic corrections of the relative order $\mathcal{O}(\alpha^2)$.
The radiative correction, beyond that included by the free electron $g$-factor, is \cite{Eides:01}
\begin{align}
H^{(6)}_{\rm rad} = Z\,\alpha^2\biggl(\ln 2-\frac{5}{2}\biggr)\,
\vec I\cdot \sum_a \frac{2}{3}\frac{Z\,\alpha\,g_{\rm N}}{m\, m_{\rm N}}\,\vec \sigma_a\,\pi\,\delta^3(r_a)
\end{align}
and the corresponding $\epsilon$ factor is $\epsilon_{\rm rad} = -0.000\,384$.
The $\mathcal{O}(\alpha^2)$ relativistic and higher order corrections are much more complicated. 
They have been calculated for the ground state of $\,^9$Be$^+$ \cite{Puchalski:09}.
Here we take this result and assume that it is proportional to the Fermi contact interaction, and obtain
$\epsilon_{\rm rel} = 0.001\,664$.  The resulting total $\epsilon$-correction is 
\begin{align} \label{16}
\epsilon =  \epsilon_{\rm Z} + \epsilon_{\rm rec} + \epsilon_{\rm rad} + \epsilon_{\rm rel} = 0.654\cdot10^{-3}.
\end{align}
Some previous works present these multiplicative corrections for all individual hyperfine contributions,
but in our opinion this cannot be fully correct because higher order relativistic corrections
may include additional terms, beyond that in $H_{\rm eff}$ in Eq.~(\ref{EHeff}). These corrections
are expected to be smaller than the experimental uncertainty and therefore are neglected here.

\section{Wave function}

In this section, we introduce the angular momentum formalism appropriate for 
explicitly correlated multielectron wave functions, i.e. represented in the basis functions 
which do not factorize into one-electron terms. This formalism accounts for 
all symmetries of the wave function present in atoms and enables
straightforward handling of the matrix elements.

\subsection{Many-electron angular factor}

The angular part of the wave function is represented in terms of the modified solid harmonics, which are adapted here for use with explicitly correlated  basis functions.
We define the solid harmonics as 
\begin{align}
{\cal Y}_{LM}(\vec r) = \sqrt{4\,\pi}\,N_L\,r^L\,Y_{LM}(\hat r)\,,
\end{align}
with some coefficients $N_L$ to be determined, involving the standard spherical harmonics 
$Y_{LM}(\theta,\phi) \equiv Y_{LM}(\hat r)$ with $\hat r = \vec r/r$.
We recall the addition theorem for spherical harmonics
 \begin{align}
 \frac{1}{2\,L+1}\sum_{M=-L}^L\, Y^*_{LM}(\hat r')\,Y_{LM}(\hat r) = \frac{1}{4\,\pi}\,P_L(\hat r'\cdot\hat r)\,,
 \end{align}
where $P_L$ are the Legendre polynomials of the order $L$. 
The corresponding formula for the solid harmonics becomes
\begin{align}\label{Esolhat}
&\frac{1}{2\,L+1}\sum_{M=-L}^L\, {\cal Y}^*_{LM}(\vec r')\,{\cal Y}_{LM}(\vec r)
 = A^2_L\,r'^L\,r^L\,P_L(\hat r'\cdot\hat r) \nonumber\\
&= (r'^{i_1}\,r'^{i_2}\,r'^{i_3}\ldots r'^{i_L})^{(L)}\,(r^{i_1}\,r^{i_2}\,r^{i_3}\ldots r^{i_L})^{(L)},
 \end{align}
where $(r^{i_1}\,r^{i_2}\,r^{i_3}\ldots r^{i_L})^{(L)}$ is a traceless and symmetric tensor 
of the order $L$ constructed from the vector $\vec r$ with Cartesian indices $i_1,i_2,i_3,\,\ldots,\,i_L$.
The last equality in Eq.~(\ref{Esolhat}) determines the factor $N_L$, which is related 
to the coefficient of $x^L$ in the Legendre polynomial $P_L(x)$, yielding
\begin{equation}
 N_L^{-2} = \frac{1}{2^L}\,{2\,L \choose L}\,.
\end{equation}
For example, $N_0 = 1$, $N_1 = 1$, $N_2 =  \sqrt{2/3}$, $N_3 =  \sqrt{2/5}$, and so on for
consecutive $L$.
 
In the correlated wave function, the total angular momentum may come from an arbitrary electron or from an arbitrary combination of many electron angular momenta.
Therefore, we introduce the following generalization of the solid harmonic
\begin{align}\label{Egsh}
&{\cal Y}_{LM}(\vec{\rho}_1,\vec{\rho}_2,\ldots,\vec{\rho}_L) \nonumber\\
&\equiv  \frac{1}{L!}\,(\vec{\rho}_1\cdot\vec\nabla_r)\,(\vec{\rho}_2\cdot\vec\nabla_r)\,\ldots
 (\vec{\rho}_L\cdot\vec\nabla_r)\, {\cal Y}_{LM}(\vec r)\,.
\end{align}
Here, $\vec{\rho}_c$ stands for either an arbitrary single electron variable $\vec{r}_a$
or for a cross product of any pair of electrons $\vec{r}_a\times\vec{r}_b$.
Note that because of the $L$-fold differentiation, the right-hand side of Eq.~(\ref{Egsh})
is $r$-independent. The function ${\cal Y}_{LM}(\vec{\rho}_1,\vec{\rho}_2,\ldots,\vec{\rho}_L)$ 
is symmetric in all its arguments and has the variable overloading property
${\cal Y}_{LM}(\vec r,\vec r,\ldots,\vec r) = {\cal Y}_{LM}(\vec r)$.
It also obeys the following summation rule
\begin{align}\label{Eshsr}
&\frac{1}{2\,L+1}\sum_{M=-L}^L\, {\cal Y}^*_{LM}(\vec{\rho}'_1, \vec{\rho}'_2,\ldots,\vec{\rho}'_L)\,{\cal Y}_{LM}(\vec{\rho}_1,\vec{\rho}_2,\ldots,\vec{\rho}_L) \nonumber\\
&= (\rho'^{i_1}_1\,\rho'^{i_2}_2\,\rho'^{i_3}_3\ldots\rho'^{i_L}_L)^{(L)}\,(\rho^{i_1}_1\,\rho^{i_2}_2\,\rho^{i_3}_3\ldots\rho^{i_L}_L)^{(L)}\,.
\end{align}
This identity allows all the matrix elements to be expressed
in terms of the scalar product, which is easy to handle with the explicitly correlated basis functions. 

Let us start with the wave function having definite orbital and spin quantum numbers $L$ and  $S$ 
and the corresponding projection quantum numbers $M_L$ and $M_S$ 
\begin{equation}
\label{PsiLSJ}
\Psi^{LSM_LM_S} = \sum_n t_n \psi^{LSM_LM_S}_n \,,
\end{equation}
where $t_n$ are linear coefficients of the expansion. 
Each basis function $\psi^{LSM_LM_S}_n$ is an antisymmetrized product of a spatial 
and spin function
\begin{equation}
\label{psiLSJ}
\psi^{LSM_LM_S} = {\cal A}\left[\phi_{L M_L}(\{ \vec r_a\})\,\chi^{S M_S}_{\{a\}}\right],
\end{equation}
where the spatial function is a product of the generalized solid harmonic 
and a function $\phi$ that depends only on interparticle distances 
\begin{equation}
\phi_{L M_L}(\{\vec r_a\})={\cal Y}_{LM_L}(\vec{\rho}_1,\vec{\rho}_2,\ldots,\vec{\rho}_L)
 \,\phi(\{\vec r_a\})\,.
\end{equation}
Let us now apply this formalism to the four-electron wave function of the beryllium atom
and write explicitly the basis functions employed for atomic levels of different symmetry.

\subsection{Four-electron basis functions}

Let $\{a\}$ and $\{\vec r_a\}$ denote a sequence of electron indices $1,2,3,4$ 
and spatial coordinates $\vec r_1, \vec r_2, \vec r_3, \vec r_4$, respectively. 
A singlet state spin wave function $\chi^{SM_S}$, for $\{a\}$ fixed at permutation $(1,2,3,4)$, 
has the form
\begin{equation}
\chi^{00} = \frac{1}{2} (\alpha_1\,\beta_2-\beta_1\,\alpha_2)\,(\alpha_3\,\beta_4-\beta_4\,\alpha_3)
\end{equation}
and the corresponding triplet state functions are
\begin{eqnarray}
\chi^{1-1} &=& \frac{1}{\sqrt{2}} (\alpha_1\,\beta_2-\beta_1\,\alpha_2)\,\beta_3\,\beta_4 \\
\chi^{10} &=& \frac{1}{2} (\alpha_1\,\beta_2-\beta_1\,\alpha_2) (\alpha_3\,\beta_4+\beta_4\,\alpha_3) \\
\chi^{11} &=& \frac{1}{\sqrt{2}} (\alpha_1\,\beta_2-\beta_1\,\alpha_2)\,\alpha_3\,\alpha_4\,.
\end{eqnarray}
In matrix elements of an arbitrary operator, all spin degrees of freedom can be 
reduced algebraically, see Sec.~\ref{Sec:Rscalar}, to a spin-free expression.
Having this in mind and the summation rule Eq. (\ref{Eshsr}), one can replace the basis functions $\phi_{L M_L}$ expressed in terms
the solid harmonics by corresponding Cartesian basis functions
\begin{equation}
\phi^{i_1\ldots i_L}=(\rho^{i_1}_1\,\rho^{i_2}_2\,\rho^{i_3}_3\ldots\rho^{i_L}_L)^{(L)}\,\phi(\{\vec r_a\}),
\end{equation}
where the variables $\rho_c$ were defined beneath Eq.~(\ref{Egsh}). 
Then, the spatial part of the basis function takes the following explicit forms:\\[1ex]
-- for $S$ states
\begin{align}
 \phi_S\equiv\phi =&\ \exp \big[-\sum_b \zeta_b \,r^2_b -\sum_{c<d} \eta_{cd} \,r^2_{cd} \big],
\label{phiS} 
\end{align}
with the nonlinear parameters $\zeta$ and $\eta$ determined variationally,\\
-- for odd $P$ states ($\vec{\rho}_1=\vec{r}_p$)
\begin{align}
\phi_P^i =&\  r^i_p\,\phi\,;
\label{phiP} 
\end{align}
-- for even $P$ states ($\vec{\rho}_1=\vec{r}_p\times\vec{r}_q$)
\begin{align}
\phi_{P^e}^{i} =&\  \epsilon^{ijk}\,r^j_p r^k_q\,\phi\,,
\label{phiPe}
\end{align}
where $\epsilon^{ijk}$ is the Levi-Civita symbol, and finally\\
-- for even $D$ states ($\vec{\rho}_1=\vec{r}_p,\vec{\rho}_2=\vec{r}_q$)
\begin{align}
\phi_D^{ij} =&\ \bigg(\frac{r_p^i r_q^j + r_p^j r_q^i}{2}  - \frac{\delta^{ij}}{3}\,r_p^k r_q^k \bigg) \,\phi\,. 
\label{phiD}
\end{align}
The subscripts $p$ and $q$ refer to arbitrary electrons (including the same ones), 
so that angular momentum may come from all the electrons in different combinations. 
A contribution to the expansion~(\ref{PsiLSJ})
from such different combinations can be optimized in a global minimization 
of the nonrelativistic energy. 

In all matrix elements, the spin part is algebraically reduced and
the angular part of the $\phi^{L M_L}$ function is converted into its Cartesian 
representation using the summation rule for solid harmonics of Eq.~(\ref{Eshsr}),
so that the final formulas can all be represented in terms of simple reduced matrix elements, 
which are convenient to use with explicitly correlated functions. 
This spin  reduction is described in the following section.

\section{Matrix elements}

The matrix element of an arbitrary operator $Q$ is 
\begin{equation}
\label{bfQ}
\langle\Psi| Q |\Psi \rangle = \sum_n\sum_m t_n^* t_m \bigl\langle \psi_n | Q | \psi_m \bigr\rangle
\end{equation}
where we skipped the angular $LSM_LM_S$ or $JM$ superscript over the wave function, because
all formulas below will be independent on the angular representation of the wave function. 
The operator $Q$ can adopt a variety of shapes according to nonrelativistic Hamiltonian and
relativistic corrections. In the operator $Q$ we can distinguish in general the spatial part $O$ 
(scalar, vector, or tensor) and its spin part involving Pauli matrices $\sigma$. 
Below, we briefly describe the reduction of the matrix elements performed
to get rid of the spin degrees of freedom. Such reduced matrix elements are assigned 
a double-braket symbol $\llangle \rrangle$.

\subsection{Reduction of the scalar matrix elements}
\label{Sec:Rscalar}

Let us start from the spin independent operator $O$, for which 
\begin{align}
\langle\psi'|O|\psi \rangle =& \hat I\,\llangle \phi'|O |\phi \rrangle \,, \label{scalar}
\end{align}
where $\hat I$ denotes the identity operator in the angular momentum subspace.
Namely, if we assume a $JM$ representation, then
\begin{align}
\langle \psi^{J'M'}|\psi^{JM}\rangle &=\langle J' M' |J M\rangle\,\llangle\phi'|\phi\rrangle.
\end{align}
The analogous formula holds for $L M_L$ and $S M_S$ representation, so Eq.~(\ref{scalar})
is independent of the angular momentum representation, as in all the formulas below in this subsection.

The reduced matrix element in  Eq. (\ref{scalar}) is defined by
\begin{align}
\llangle \phi' | O | \phi \rrangle &= \bigl\langle \phi'(\{\vec r_a\})|O\, {\cal A}
             [u_l\,\phi(\{ \vec r_b\})] \bigr\rangle .
\end{align}
In the above expression ${\cal A}$ denotes the sum over all $n!$ permutations of $n$ electrons
\begin{equation}
\mathcal{A}=\sum_{l=1}^{n!}\varepsilon_l\,\mathcal{P}_l\,.
\end{equation}
The coefficients
\begin{equation}
u_l=\varepsilon_l\langle\chi'\,|\,\hat{\Xi}\,\mathcal{P}_l\,\chi\rangle,\text{ with }
\Xi=1,\,\vec{\sigma}_a,\text{ or  }\,\vec{\sigma}_a\,\vec{\sigma}_b\,,
\end{equation}
which accompany the right function $\phi$, depend on particular permutation
$\mathcal{P}_l$ and are explicitly shown for $n=4$ in Table~\ref{u-coefficients} of Appendix~\ref{App:u-coeff}. 
The reduced matrix element may have implicit summation over Cartesian indices; then,
$\llangle \phi'|O |\phi\rrangle$ denotes $\llangle \phi'^i|O |\phi^i \rrangle$ 
or   $\llangle \phi'^{ij}O |\phi^{ij} \rrangle$ depending on the angular momentum of the state in question.
We will skip these Cartesian indices as long as it does not lead to any confusion.
These reduced matrix elements are a workhorse of this approach. For example, 
the matrix elements of the nonrelativistic Hamiltonian $H$ can also be expressed in terms 
of the reduced ones. The fact that we originally did not us the wave function with specified 
$J$ and $M$ is irrelevant. The nonrelativistic Hamiltonian $H$  does not depend on $J$ or $M$, 
so different $\psi^{LSJM}$ will lead to the same matrix elements as long as $L$ and $S$ are fixed.
             
\subsection{Reduction of spin-dependent operators}

Similarly, all the matrix elements of the spin-dependent operators in Eq.~(\ref{EHfs}) 
can be expressed in terms of the reduced ones as follows
\begin{align}
 \langle \psi' |\sum_a \vec \sigma_a \cdot \vec O_a |\psi \rangle &=   
- \frac{\vec L\cdot \vec S}{(L+1)}\, \mathrm{i} \,\epsilon^{ijk} \sum_a\llangle{\phi'}^{i} | O_a^j | \phi^{k} \rrangle_{a}\,, \label{Els}\\
 \langle \psi' |\sum_{a<b} \sigma^i_a\,\sigma^j_b\,O^{ij}_{ab} |\psi \rangle &=   
\frac{12\,(L^i\,L^j)^{(2)}\,(S^i\,S^j)^{(2)}}{(2\,L+3)\,(L+1)}\nonumber \\ 
&\quad\times\sum_{a<b}\llangle{\phi'}^{i} | O_{ab}^{ij} | \phi^{j} \rrangle_{ab}\,, \label{Els2}
\end{align}
where $\vec S = \frac{1}{2}\sum_a \vec\sigma_a$.
Again, the angular indices of the wave function $\psi$ are skipped because this angular part 
goes to matrix elements of $\vec L\cdot\vec S$ or $(L^i\,L^j)^{(2)}\,(S^i\,S^j)^{(2)}$ operators. 

The above reduced matrix elements are defined as 
\begin{align}
\llangle\phi' | O_c | \phi \rrangle_{c} &=   
\bigl\langle\phi'(\{\vec r_a\})|
O_c\,{\cal A} \big[u^{c}_{l}\,\phi(\{\vec r_b\})\big] \big \rangle\,,\\
\llangle\phi' | O_{cd} | \phi \rrangle_{cd} &=   
\bigl\langle\phi'(\{\vec r_a\})|
O_{cd}\,{\cal A} \big[u^{cd}_{l}\,\phi(\{\vec r_b\})\big] \big \rangle\,,
\end{align}
and have the advantage that they involve only scalar operators built of spatial variables $\vec r_a$,
and therefore they can all easily be evaluated in an explicitly correlated basis,
in particular in the ECG basis. Moreover, these reduced matrix elements have the following properties
\begin{align}
\sum_c\llangle\phi' | O | \phi \rrangle_{c} &= 2\,\llangle\phi' | O | \phi \rrangle\,, \label{prop1}\\
\sum_{c<d}\llangle\phi' | O | \phi \rrangle_{cd} &= -\llangle\phi' | O | \phi \rrangle\,, \label{prop2}
\end{align}
which will be used  to prove the above reduction formulas. For these proofs we shall need also 
the following two equalities
\begin{align}
-\mathrm{i}\,\epsilon^{ijk}\llangle\phi'^i|L^j|\phi^k\rrangle&=(L+1)\,\llangle\phi'|\phi\rrangle\,, \label{identt1}\\
\llangle\phi'^i|(L^i\,L^j)^{(2)}|\phi^k\rrangle&=-\frac{1}{6}\,(L+1)\,(2\,L+3)\,\llangle\phi'|\phi\rrangle\,. \label{identt2}
\end{align}
The reduction formulas are independent of the operator $\vec O_a$. So, to prove Eq.~(\ref{Els}) 
let $\vec O_a = \vec L$ be the orbital angular momentum, then 
\begin{align}
l.h.s. =&\ \langle \psi' |\sum_a \vec \sigma_a \cdot \vec L |\psi \rangle = \langle \psi' |2\,\vec S\,\vec L\,|\psi \rangle = 2\,\vec S\,\vec L\,\llangle \phi' |\phi \rrangle\,.
\end{align}
Using Eqs. (\ref{prop1}) and (\ref{identt1}), the right hand side of Eq.~(\ref{Els}) can be rearranged to
\begin{align}
r.h.s=&\ - \frac{\vec L\cdot \vec S}{L+1}\, \mathrm{i} \,\epsilon^{ijk} \sum_a\llangle{\phi'}^{i} | L^j | \phi^{k} \rrangle_{a} 
\nonumber \\ =&\ 
\frac{2}{L+1}\,\vec L\cdot \vec S\, (-\mathrm{i}) \,\epsilon^{ijk} \llangle{\phi'}^{i} | L^j | \phi^{k} \rrangle
= 2\,\vec S\,\vec L\,\llangle \phi' |\phi \rrangle\,,
\end{align}
which is equal to $l.h.s$. Similarly, to prove Eq.~(\ref{Els2}), let $O_{ab}^{ij} = (L^i\,L^j)^{(2)}$;
then
 \begin{align}
l.h.s. &=\langle \psi' |\sum_{a<b} \sigma^i_a\,\sigma^j_b\,(L^i\,L^j)^{(2)} |\psi \rangle 
 \nonumber \\ &=
  2\, \langle \psi' | S^i\,S^j\,(L^i\,L^j)^{(2)} |\psi \rangle 
  \nonumber \\ &=
 2\, (S^i\,S^j)^{(2)}\,(L^i\,L^j)^{(2)}\,\llangle\phi' |\phi \rrangle\,.
 \end{align}
Taking Eq.~(\ref{identt2}), the right hand side of Eq.~(\ref{Els2}) becomes
 \begin{align}
r.h.s. =&\  -\frac{12\,(L^i\,L^j)^{(2)}\,(S^i\,S^j)^{(2)}}{(2\,L+3)(L+1)}\,\llangle{\phi'}^{i} | (L^i\,L^j)^{(2)} | \phi^{j} \rrangle
 \nonumber \\ =&\ 
2\,(L^i\,L^j)^{(2)}\,(S^i\,S^j)^{(2)}\,\llangle\phi' | \phi\rrangle\,,
 \end{align}
which is equal to $l.h.s$.

\subsection{Reduction of the vector and tensor matrix elements}

Analogous reductions can be performed for the hyperfine operators in $H_{\rm hfs}$, namely
\begin{align}
\langle \psi' |\sum_a \vec \sigma_a \, O_a |\psi \rangle
&=\vec S \,\sum_a\llangle\phi' | O_a | \phi \rrangle_{a}\,,\\
\langle \psi' | \vec O |\psi \rangle  &= 
 -\frac{\vec L}{(L+1)}\,\mathrm{i}\,\epsilon^{ijk} \llangle\phi^i |O^j | \phi^k \rrangle\,, \\
\langle \psi' |\sum_a \sigma_a^j \,O_a^{ij} | \psi \rangle &= 
\frac{-6\,S^j\,(L^i\,L^j)^{(2)}}{(2\,L+3)\,(L+1)}\,\sum_a\llangle\phi^i|O_a^{ij}|\phi^j\rrangle_{a}\,,\\
\langle \psi' | O^{ij} | \psi \rangle &=
\frac{-6\,(L^i\,L^j)^{(2)}}{(2\,L+3)\,(L+1)}\,\bigl\llangle\phi^i | O^{ij} | \phi^j \big \rrangle\,.
\end{align}
The proofs of the above reduction formulas are very similar to those shown in the preceding subsection.
One assumes that $O_a = \hat I$, $\vec O = \vec L$, $O_{ab}^{ij} = (L^i\,L^j)^{(2)}$,
and repeats the previous proofs correspondingly.

\section{Effective fine/hyperfine Hamiltonian}
In order to account for the combined fine and hyperfine structure with 
an arbitrary coupling strength, it is necessary to extend the original formulation 
of the hyperfine splitting theory by Hibbert \cite{Hibbert:75} and represent the fine 
and the hyperfine structure of an arbitrary state in terms of an effective Hamiltonian,
instead of expectation values. The effective Hamiltonian reads 
\begin{align}\label{EHeff}
H_\mathrm{eff}&=c_0+c_1\,\vec{L}\cdot\vec{S} +c_2\,(L^iL^j)^{(2)}(S^iS^j)^{(2)}\nonumber \\ 
&\quad+a_1\,\vec{I}\cdot\vec{S}+a_2\,\vec{I}\cdot\vec{L}+a_3\,(L^iL^j)^{(2)}S^iI^j \nonumber \\ 
&\quad+\frac{b}{6}\,\frac{3\,(I^i\,I^j)^{(2)}}{I\,(2\,I-1)}\,\frac{3\,(L^iL^j)^{(2)}}{L\,(2\,L-1)},
\end{align}
where the coefficients $a_1, a_2, a_3, b, c_0, c_1$, and $c_2$ are independent of $\vec J= \vec L+\vec S$
but are specific to the particular state. The $c_0$ coefficient is the so-called centroid energy, 
which in our case is 
\begin{align}
c_0 = E_0+E_{\rm ns}^{(4)} + \mathcal{O}(\alpha^5),
\end{align}
where $E_{\rm ns}^{(4)} = \langle H_{\rm ns}^{(4)} \rangle$ is the spin-independent relativistic
correction. This correction can be rewritten as
\begin{equation}
E_{\rm ns}^{(4)} = -\frac{1}{8} V_1 + \frac{Z}{8} V_2 + \frac{1}{4} V_3 -\frac{1}{2} V_4  +\frac{Z}{2\,m_\mathrm{N}} V_5 \,,
\end{equation}
with $V_1,\dots, V_5$ defined in Table~\ref{tblredmel1}.
The fine structure parameters $c_1$ and $c_2$, using formulas from the previous section, are
\begin{align}
c_1 =&\ -\frac{1}{(L+1)}\, \bigg[\frac{Z}{4} \Big( (g-1)V_{\mathrm{f}1} - \frac{g}{m_\mathrm{N}} V_{\mathrm{f}4}\Big) \nonumber \\
& + \frac{1}{4} \Big(g\,V_{\mathrm{f}2}  -  (g-1) V_{\mathrm{f}3} \Big) \bigg]\,,\\
c_2 =&\ - \frac{12}{(2\,L+3)(L+1)}\,\frac{3\,g^2}{16} V_{\mathrm{f}5} \,,
\end{align}
while the hyperfine structure parameters are
\begin{align}
a_1 &=\frac{Z}{m_\mathrm{N}}\,\frac{g\,g_\mathrm{N}}{12}\,V_{\mathrm{h}1}\,,\\
a_2 &=-\frac{1}{(L+1)}\,\frac{Z}{m_\mathrm{N}}\,\bigg(\frac{g_\mathrm{N}}{2}\,V_{\mathrm{h}2} - \frac{g_\mathrm{N}-1}{2\,m_\mathrm{N}}\,V_{\mathrm{h}3}\bigg),\\
a_3 &=\frac{-6}{(2\,L+3)\,(L+1)}\,\frac{Z}{m_\mathrm{N}}\,\frac{3\,g\,g_\mathrm{N}}{8}\,V_{\mathrm{h}4}\,,\\ \nonumber\\
b   &= \frac{6\,L\,(2\,L-1)}{(2\,L+3)\,(L+1)}\,Q_\mathrm{N}\,V_{\mathrm{h}5}\,.
\end{align}
The expectation values $V_{\mathrm{f}i}$ and $V_{\mathrm{h}i}$ used to determine the fine and hyperfine
parameters are defined in Table~\ref{tblredmel2}.
Once these parameters are calculated, 
the effective hyperfine structure Hamiltonian $H_\mathrm{eff}$ can be diagonalized, for example in 
the $|L,M_L;S,M_S;I,M_I\rangle$ basis, yielding the combined fine/hyperfine levels with respect 
to the centroid energy $c_0$.

\section{Calculations and results}

\subsection{Centroid energies}

\subsubsection{Variational optimization of the nonrelativistic energy}
In the numerical calculations we followed closely our previous works 
devoted to the singlet $S$ and $P$ states of beryllium \cite{Puchalski:13a, Puchalski:14a}.  
We used the wave functions expanded in the basis of ECG functions~(\ref{phiS})-(\ref{phiD}), 
whose non-linear parameters were variationally optimized.
The optimization was performed at the infinite nuclear mass limit of the nonrelativistic Hamiltonian,
Eq.~(\ref{H2}). Then, the nonrelativistic energies and the wave functions of $^9$Be were 
generated with the same nonlinear parameters without significant lose of accuracy. 
In order to achieve numerical accuracy \hbox{$\sim\!10^{-9}$} for nonrelativistic energy $E_0$, 
which is equivalent to a numerical accuracy of energy levels $<0.01\,\icm$,  
we assumed the maximum size of the basis sets equal to 4096, 6144, and 8192
for $S$-, $P$-, and $D$-states, respectively. A sequence of energies obtained for consecutive basis sets enabled extrapolation to
the complete basis limit and estimation of the error resulting from basis set truncation.
The nonrelativistic energy $E_0$ convergence for all the studied states of $^\infty$Be is presented 
in Table~\ref{Tconv}. Note that the rate of the convergence depends on the given atomic state.

\begin{table*}[!hbt]
\renewcommand{\arraystretch}{1.2}
\caption{Convergence of the nonrelativistic energy $E_0$ of $^\infty$Be (in a.u.) and comparison
with other ECG results, or if not available, with the most accurate value from another method.}
\label{Tconv}
\begin{ruledtabular}
\begin{tabular}{rw{5.14}rw{5.14}rw{5.14}}
\centt{Size}       &  \cent{2s3s\,^3\!S}  & \centt{Size} & \cent{2s2p\,^1\!P} & \centt{Size} & \cent{2s2p\,^3\!P} \\
\hline
 768               & -14.430\,065\,800\,88  &     1024 & -14.473\,445\,215\,92 &     1024 & -14.567\,241\,485\,35 \\
1024               & -14.430\,066\,834\,12  &     1536 & -14.473\,449\,010\,68 &     1536 & -14.567\,243\,359\,33 \\
1536               & -14.430\,067\,459\,90  &     2048 & -14.473\,450\,455\,64 &     2048 & -14.567\,243\,913\,64 \\
2048               & -14.430\,067\,579\,18  &     3072 & -14.473\,451\,162\,07 &     3072 & -14.567\,244\,114\,67 \\
3072               & -14.430\,067\,637\,28  &     4096 & -14.473\,451\,310\,50 &     4096 & -14.567\,244\,192\,08 \\
4096               & -14.430\,067\,666\,35  &     6144 & -14.473\,451\,361\,77 &     6144 & -14.567\,244\,215\,84 \\
$\infty$           & -14.430\,067\,678(7)   & $\infty$ & -14.473\,451\,384(9) & $\infty$ & -14.567\,244\,232(8)  \\
\cite{Frolov:09} 7000
                   & -14.430\,059\,43       & \cite{Stanke:19} 16400
                                                        & -14.473\,451\,388\,2  & \cite{Kedziorski:20} 8000
                                                                                            &  -14.567\,244\,222 \\[0ex]
                                                                                            \hline
\centt{Size}      & \cent{2p^2\,^3\!P^e} & \centt{Size} & \cent{2p^2\,^1\!D}  & \centt{Size} & \cent{2s3d\,^3\!D}   \\
\hline
1024              & -14.395\,452\,640\,71 & 1536     & -14.408\,232\,496\,49 & 1536     & -14.384\,631\,192\,32 \\
1536              & -14.395\,453\,441\,97 & 2048     & -14.408\,234\,916\,82 & 2048     & -14.384\,632\,963\,77 \\
2048              & -14.395\,453\,625\,95 & 3072     & -14.408\,236\,788\,28 & 3072     & -14.384\,633\,859\,38 \\
3072              & -14.395\,453\,700\,13 & 4096     & -14.408\,237\,032\,51 & 4096     & -14.384\,634\,414\,54 \\
4096              & -14.395\,453\,720\,27 & 6144     & -14.408\,237\,213\,69 & 6144     & -14.384\,634\,572\,57 \\
6144              & -14.395\,453\,738\,26 & 8192     & -14.408\,237\,270\,12 & 8192     & -14.384\,634\,603\,77 \\
$\infty$          & -14.395\,453\,745(4)  & $\infty$ & -14.408\,237\,290(9)  & $\infty$ & -14.384\,634\,616(7)   \\
\cite{Zhu:95} FCPC& -14.395\,431\,6       & \cite{Stanke:19b} 12300
                                                     & -14.408\,237\,282 & \cite{Sharkey:14} 8100
                                                                                        & -14.384\,634\,597\,13 
\end{tabular}
\end{ruledtabular}
\end{table*}

This table contains also the best currently available literature results. 
For the $2s3s\,^3\!S$ state the energy reported by Frolov and Wardlaw \cite{Frolov:09} 
seems to be rather poorly converged---despite using a 7000-term ECG expansion their result
is about $8\cdot 10^{-6}\text{ a.u.}\approx 2\,\icm$ above our energy obtained with 4096-term
wave function. Significantly longer ECG expansions have been employed 
for the $2s2p\,^1\!P$, $2s2p\,^3\!P$, and $2p^2\,^1\!D$ states by Adamowicz {\em et al.}
\cite{Stanke:19,Kedziorski:20,Stanke:19b}. In these cases, their variational energy
is by $10^{-8}-10^{-9}$ a.u. lower than our upper bound,
whereas for the $2s3d\,^3\!D$ state our upper bound slightly improves over 
the variational energy obtained by Sharkey \etal\ \cite{Sharkey:14} 
from an equivalent ECG expansion. The best previous calculations 
of the nonrelativistic energy for the $2p^2\,^3\!P$ state were obtained using 
a full-core plus correlation (FCPC) method \cite{Zhu:95} and gave the energy 
almost $5\,\icm$ higher than the current one.

In general, the current state-of-the-art calculations offer a relative accuracy 
of the order of $10^{-10}$, which corresponds to $\approx\!10^{-4}\,\icm$ of absolute accuracy.
Still, there seems to be room for further accuracy improvement of the ECG method
in relation to four-electron atoms, either by increasing the basis size or by tuning 
the optimization algorithms. However, the ability to maintain reliable numerical convergence 
is limited due to the double precision arithmetic used in the algorithms. 
Significant improvement of the current results will require the use 
of higher precision arithmetic and bases of size \hbox{$>20\,000$}, which means 
a dramatic increase in the computation time. This suggests the need to redesign current 
ECG algorithms or look for new, more efficient solutions in the future.

\subsubsection{Calculations of reduced matrix elements}

The finite-mass wave functions were subsequently employed in the  evaluation of matrix elements.
The values of all the reduced matrix elements for relativistic and QED corrections 
along with the nonrelativistic energy and the Bethe logarithm are collected
in Tables~\ref{tblredmel1},~\ref{tblredmel2}. All the entries represent extrapolated values
with estimated uncertainty. Because the use of original formulas for singular operators 
leads to a slow numerical convergence (this spurious effect is particularly exposed
in calculations using Gaussian-type basis functions having improper short-distance 
behavior), regularized versions of matrix elements were employed following the rules 
provided in Appendix~\ref{App:regularization}.
For the $2s2p\,^1\!P$ state the Bethe logarithm, $\ln k_0$, was calculated directly in 
Ref.~\onlinecite{Puchalski:13a}, and in this case the overall uncertainty is dominated 
by the higher order corrections. 
This numerical value of $\ln k_0$ was adopted also for the remaining states with
a relevantly large uncertainty assigned. Eventually, this uncertainty dominated
the overall theoretical uncertainty.
The centroid energies evaluated with these matrix elements
are put together in Table~\ref{Tcentr}.

\newcommand{\Vone}{$\sum_a \llangle\vec p^{\,4}_a\rrangle$}
\newcommand{\Vtwo}{$\sum_a \llangle 4\,\pi\,\delta^3(r_{a})\rrangle$}
\newcommand{\Vthree}{$\sum_{a<b} \llangle 4\,\pi\,\delta^3(r_{ab})\rrangle$}
\newcommand{\Vfour}{$\sum_{a<b} \llangle p_a^i\,\bigl(\frac{\delta^{ij}}{r_{ab}}+\frac{r^i_{ab}\,r^j_{ab}}{r^3_{ab}} \bigr)\, p_b^j \rrangle$}
\newcommand{\Vfive}{$\sum_a \llangle p_a^i\, \bigl(\frac{\delta^{ij}}{r_{a}}+\frac{r^i_{a}\,r^j_{a}}{r^3_{a}} \bigr)\, p_\mathrm{N}^j\rrangle$}
\newcommand{\PAS}{$\sum_{a<b} \llangle P(r_{ab}^{-3})\rrangle$}
\newcommand{\ie}{\mathrm{i}\,\epsilon^{ijk}}
\newcommand{\sa}{\sum_a}
\newcommand{\Fone}{$\ie\sa\llangle i|\bigl(\frac{\vec{r}_a}{r_a^3}\times\vec{p}_a\bigr)^j|k \rrangle_a$}
\newcommand{\Ftwo}{$\ie\sa\sum_{b\neq a}\llangle i|\bigl(\frac{\vec{r}_{ab}}{r_{ab}^3}\times\vec{p}_b\bigr)^j|k\rrangle_a$}
\newcommand{\Fthree}{$\ie\sa\sum_{b\neq a}\llangle i|\bigl(\frac{\vec{r}_{ab}}{r_{ab}^3}\times\vec{p}_a\bigr)^j|k\rrangle_a$}
\newcommand{\Ffour}{$\ie\sa\llangle i|\bigl(\frac{\vec{r}_a}{r_a^3}\times\vec{p}_\mathrm{N}\bigr)^j|k \rrangle_a$}
\newcommand{\Ffive}{$\sum_{a<b}\llangle i|\frac{r_{ab}^i\,r_{ab}^j}{r_{ab}^5}-\frac{\delta^{ij}}{3\,r_{ab}^3}|j\rrangle_{ab}$}
\newcommand{\Hone}{$\sa\llangle 4\,\pi\,\delta^3(r_{a})\rrangle_a$}
\newcommand{\Htwo}{$\ie\sa\llangle i|\bigl(\frac{\vec{r}_a}{r_a^3}\times\vec{p}_a\bigr)^j|k \rrangle$}
\newcommand{\Hthree}{$\ie\sa\llangle i|\bigl(\frac{\vec{r}_a}{r_a^3}\times\vec p_{\rm N}\bigr)^j|k \rrangle$}
\newcommand{\Hfour}{$\sa\llangle i|\frac{r_{a}^i\,r_{a}^j}{r_{a}^5} - \frac{\delta^{ij}}{3\,r_{a}^3}|j \rrangle_a$}
\newcommand{\Hfive}{$\sa\llangle i|\frac{r_{a}^i\,r_{a}^j}{r_{a}^5} - \frac{\delta^{ij}}{3\,r_{a}^3}|j\rrangle$}
\begin{table*}[!hbt]
\renewcommand{\arraystretch}{1.2}
\caption{Expectation values of various operators and spin-independent reduced matrix elements 
of $^9$Be (in a.u.).}
\label{tblredmel1}
\begin{ruledtabular}
\begin{tabular}{lw{6.12}w{5.11}w{5.16}w{5.16}}
Reduced matrix element & \cent{2s3s\,^3\!S}   & \cent{2s2p\,^1\!P}  & \cent{2p^2\,^1\!D}  \\
\hline                                                                               
$E_0$            & -14.429\,160\,970(6) & -14.472\,543\,762(11)& -14.407\,351\,381(10) \\
$V_1=$   \Vone   & 2\,149.48\,58(8)     & 2\,132.787\,7(18)    & 2\,109.009\,3(9)    \\
$V_2=$   \Vtwo   & 441.618\,01(6)       & 438.458\,40(4)       & 434.113\,91(8)      \\
$V_3=$   \Vthree & 19.904\,846(4)       & 19.700\,026(7)       & 19.370\,006(5)      \\
$V_4=$   \Vfour  & 1.814\,877(4)        & 1.622\,713(4)        & 1.380\,895\,6(5)    \\
$V_5=$   \Vfive  & -223.060\,24(2)      & -220.643\,672(2)     & -217.444\,43(3)     \\
$\ln k_0$        &  5.752(3)^a          &  5.752\,32(8)^a      &  5.752(3)^a         \\
         \PAS    & -7.486\,93(2)^b      & -7.097\,17(3)^{b}    & -6.918\,15(2)^b         \\
$V_{\mathrm{h}1}=$ \Hone   & 13.294\,900(10)  &                     &                     \\
$V_{\mathrm{h}2}=$ \Htwo   &                  & -0.370\,894(3)      & -0.559\,374\,9(18)  \\
$V_{\mathrm{h}3}=$ \Hthree &                  & -0.112\,512(9)      & -0.216\,094(3)  \\
$V_{\mathrm{h}5}=$ \Hfive  &                  &  0.112\,086\,0(11)  & 0.080\,872\,8(12)   \\
\end{tabular}
\begin{flushleft}
$^a$ Adopted from $2s2p\,^1\!P$ state~\cite{Puchalski:13a};
$^b$ Calculated in the infinite mass limit.
$^c$ \cite{Puchalski:13a};
\end{flushleft}
\end{ruledtabular}
\end{table*}

\begin{table*}[!hbt]
\renewcommand{\arraystretch}{1.2}
\caption{Expectation values of various operators and spin-independent reduced matrix elements
of $^9$Be (in a.u.).}
\label{tblredmel2}
\begin{ruledtabular}
\begin{tabular}{lw{6.12}w{5.11}w{5.16}w{5.16}}
Reduced matrix element & \cent{2s2p\,^3\!P}   & \cent{2p^2\,^3\!P^{e}} & \cent{2s3d\,^3\!D}    \\
\hline                                                          
$E_0 $           & -14.566\,341\,475(4) & -14.394\,568\,519(5) & -14.383\,731\,170(6) \\
$V_1=$   \Vone   & 2\,131.397\,1(15)    & 2\,086.304\,2(12)    & 2\,144.482\,7(7)    \\
$V_2=$   \Vtwo   & 438.127\,79(11)      & 429.777\,10(13)      & 440.786\,59(9)        \\
$V_3=$   \Vthree & 19.684\,698(3)       & 19.065\,685(2)       & 19.837\,514(2)        \\
$V_4=$   \Vfour  & 1.457\,377\,9(16)    & 1.070\,998\,2(13)    & 1.809\,602\,7(12)          \\
$V_5=$   \Vfive  & -220.376\,91(6)      & -214.225\,93(5)      & -222.412\,27(6)       \\
$\ln k_0$        &  5.752(3)^a          &   5.752(3)^a         &  5.752(3)^a           \\
         \PAS    & -6.966\,49(3)^b      & -6.505\,54(4)^b      &  -7.493\,48(2)^b         \\
$V_{\mathrm{f}1}=$ \Fone   & -0.605\,451(3)       & -0.620\,435\,4(7)    & -0.016\,199\,88(6)    \\
$V_{\mathrm{f}2}=$ \Ftwo   & 0.273\,149\,3(4)     &  0.296\,033\,6(2)    & 0.010\,830\,7(4)      \\
$V_{\mathrm{f}3}=$ \Fthree & -1.124\,565(4)       &  -1.154\,002\,9(3)   &  -0.040\,076\,2(5)    \\
$V_{\mathrm{f}4}=$ \Ffour  & -0.243\,575\,3(2)    & -0.187\,062\,4(12)   & 0.003\,014\,1(4)      \\
$V_{\mathrm{f}5}=$ \Ffive  & -0.017\,082\,4(3)    & 0.012\,571\,98(2)    & -0.000\,828\,213(2)   \\
$V_{\mathrm{h}1}=$ \Hone   & 9.247\,623(18)       & -1.389\,667(5)       & 12.130\,05(3)         \\
$V_{\mathrm{h}2}=$ \Htwo   & -0.606\,202(3)       & -0.621\,281\,1(7)    & -0.016\,209\,10(9)    \\
$V_{\mathrm{h}3}=$ \Hthree & -0.240\,311(8)       & -0.237\,269(4)       &  0.003\,328\,5(8)     \\
$V_{\mathrm{h}4}=$ \Hfour  & 0.219\,150\,1(6)     & -0.218\,683\,3(9)    & 0.002\,441(2)         \\
$V_{\mathrm{h}5}=$ \Hfive  & 0.192\,574\,73(18)   & -0.194\,530\,5(4)    & 0.002\,083(2)         \\
\end{tabular}
\begin{flushleft}
$^a$ Adopted from $2s2p\,^1\!P$ state~\cite{Puchalski:13a};
$^b$ Calculated in the infinite mass limit.
\end{flushleft}
\end{ruledtabular}
\end{table*}

\subsection{Combined fine/hypefine structure}
\label{sec:fshfs}

\begin{figure}[!ht]
\includegraphics[scale=0.43]{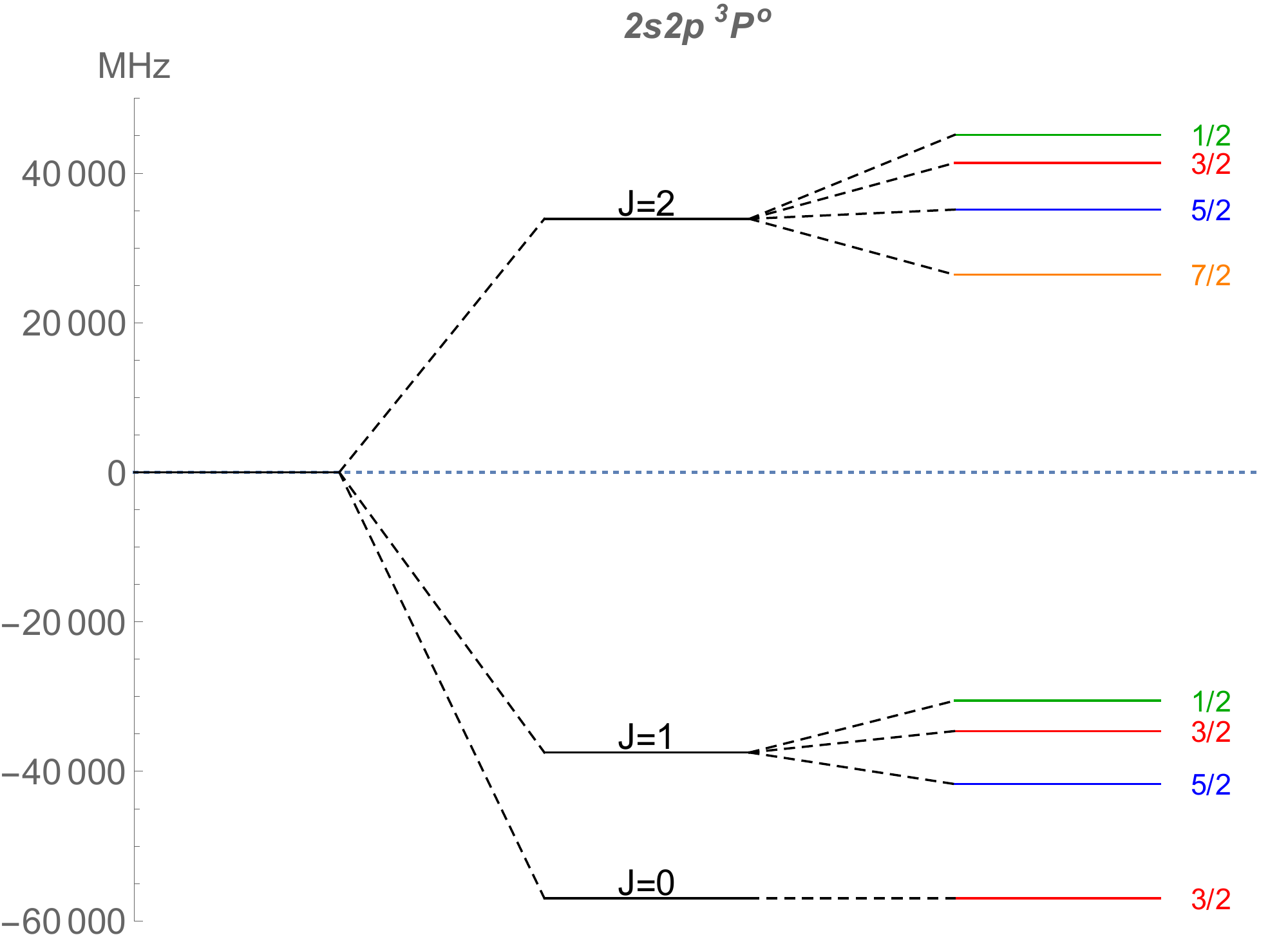}
\caption{The fine and hyperfine splitting for the $2s2p\,^3\!P^o$ state. 
The hf splitting was scaled by a factor of 20.}
\label{F3Po}
\end{figure}

\begin{figure}[!ht]
\includegraphics[scale=0.44]{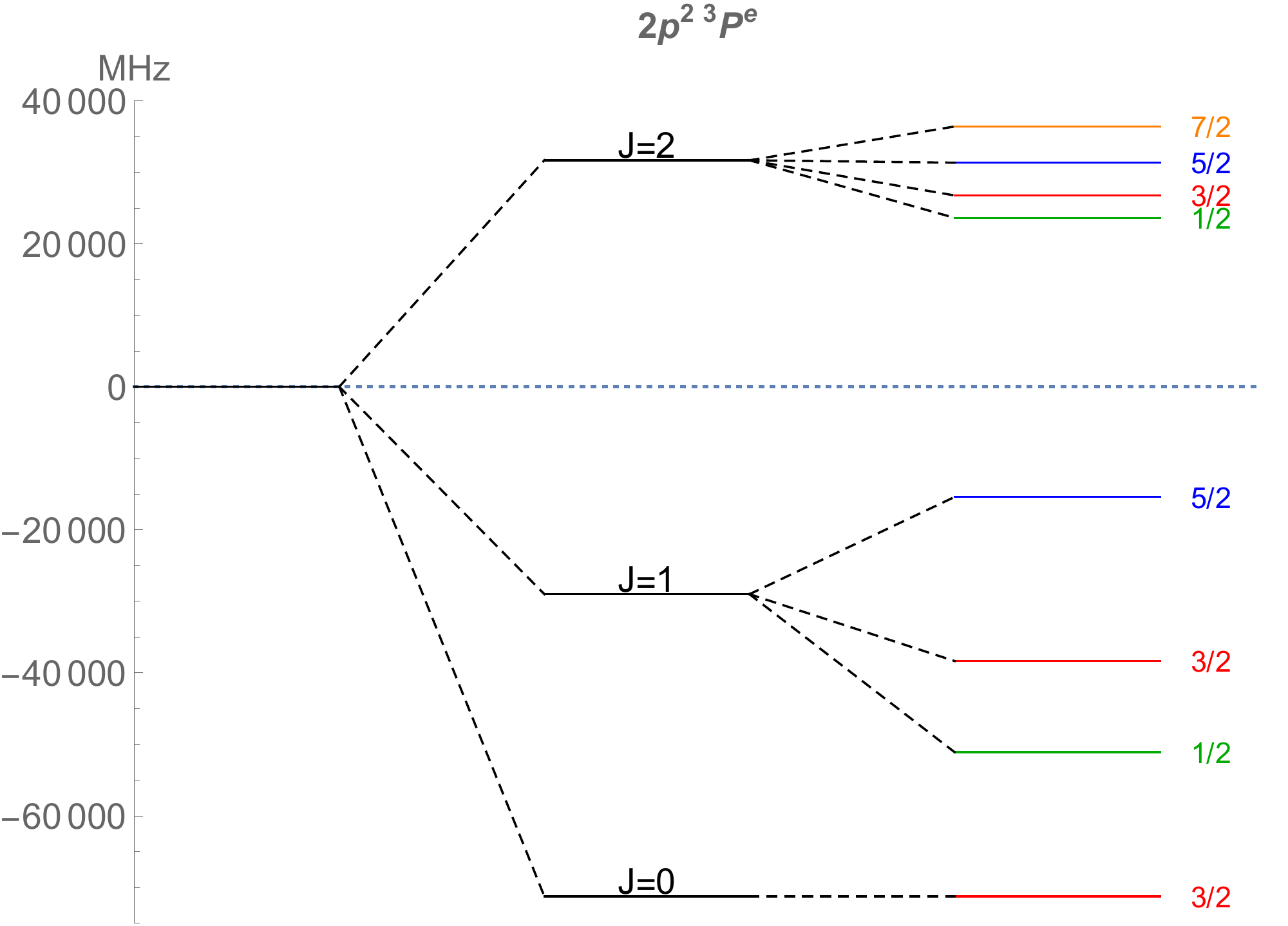}
\caption{The fine and hyperfine splitting for the $2p^2\,^3\!P^e$ state. 
The hf splitting was scaled by a factor of 500.}
\label{F3Pe}
\end{figure}

The effective Hamiltonians of the general form given by Eq.~(\ref{EHeff}) were constructed 
separately for each atomic state.
They differ from each other in numerical values of parameters $a_1$, $a_2$, $a_3$, $b$, $c_1$, and $c_2$,
listed in Table~\ref{Tabcparms}, and hence also in the number of terms included. 
These Hamiltonians were diagonalized in the basis of $|L,M_L;S,M_S;I,M_I\rangle$ states 
using standard angular momentum algebra. 
Numerical eigenvalues representing the shift of the atomic hyperfine level with respect to 
the corresponding centroid are presented in Table~\ref{Thfslevels}.
Depending on the atomic state, these hyperfine levels extend in the range from tens 
up to almost a hundred thousand MHz. Figures~\ref{F3Po}-\ref{F3D} show graphically the fine/hyperfine
splitting in the case of three angular momenta states.
The corresponding eigenfunctions, in turn, can be employed to provide intensities of transitions
between individual hyperfine levels and help to overcome the line-shape-related limitations
to the precision of contemporary measurements~\cite{Cook:18,Cook:20}.

Because the states $^3\!S_1$, $^1\!P_1$, and $^1\!D_2$ involve only two angular momenta, one can employ 
the commonly used $A_J$ and $B_J$ coefficients to represent their hyperfine structure
\begin{align}
\langle H_{\rm hfs} \rangle_J &= A_J\,\vec I\cdot\vec J + \frac{B_J}{6}\,
\frac{3\,(I^i\,I^j)^{(2)}}{I\,(2\,I-1)}\,\frac{3\,(J^i\,J^j)^{(2)}}{J\,(2\,J-1)},
\end{align}
where $\vec J$ is the total electronic angular momentum.
So, in the case of the $2s3s\,^3\!S_1$ state, $\vec J=\vec S$ and 
\begin{align}
A_1(^3\!S) &= a_1\,(1+\epsilon) = -332.50(2) \text{ MHz} \label{epsilon}\\
B_1(^3\!S) &= 0\,,
\end{align}
where $\epsilon=0.654\times 10^{-3}$ was taken from Ref.~\cite{Puchalski:21}.
In the case of the $2s2p\,^1\!P_1$ state, $\vec J=\vec L$ and
\begin{align}
A_1(^1\!P) &= a_2 = -13.888\,2(7) \text{ MHz} \\
B_1(^1\!P) &= b = 0.845\,40(4) \text{ MHz} \,.
\end{align}
Finally for the $2p^2\,^1\!D_2$ state, $\vec J= \vec L$ and
\begin{align}
A_2(^1\!D) &= a_2 = -13.964\,0(7) \text{ MHz} \\
B_2(^1\!D) &= b = 1.742\,80(9) \text{ MHz}\,.
\end{align}
   
The calculation of the fine/hyperfine structure for the $2s2p\,^3\!P$, $2p^2\,^3\!P^e$, 
and $2s3d\,^3\!D$ states requires diagonalization of the effective fine/hyperfine 
Hamiltonian in Eq.~(\ref{EHeff}). 
For the $2s3d\,^3\!D$ state, the diagonalization reveals that the $a_1$ parameter
is around three times larger than the parameters $c_1$ and $c_2$. This makes the interaction
of electronic and nuclear spins the dominating one and disqualifies $J$ as a good quantum number.
Therefore, one cannot use $A_J$ and $B_J$ coefficients---instead we present actual 
values of the fine/hyperfine levels. In addition, to account for the leading relativistic 
and radiative corrections, the $a_1$ parameter is rescaled by the $(1+\epsilon)$ factor, 
see Eq.~(\ref{epsilon}) and the related discussion in Sec.~\ref{sec:epsilon}.

\begin{table*}[!hbt]
\renewcommand{\arraystretch}{1.0}
\caption{Centroid energy contributions (in $\icm$) relative to the ground $2s^2\,^1\!S$ state  
for $^9$Be and comparison with experimental and other theoretical data. The total energy
in terms of the ionization potential is also shown at the bottom of the table.}
\label{Tcentr}
\begin{ruledtabular}
\begin{tabular}{lw{3.8}w{3.8}w{3.8}w{3.8}w{3.8}w{3.8}}
& \cent{2s2p\,^3\!P} & \cent{2s2p\,^1\!P}& \cent{2s3s\,^3\!S} & \cent{2p^2\,^1\!D} & \cent{2p^2\,^3\!P^e} & \cent{2s3d\,^3\!D} \\
\hline
$m \alpha^2$      & 21\,968.103(3) & 42\,554.325(6)  & 52\,075.750(10)& 56\,862.395(5) & 59\,667.912(2) & 62\,046.433(3)  \\
$m \alpha^4$      &      13.189(3) &      12.171(3)  &       5.726(4) &      21.968(4) &      30.728(4) &       8. 000(5)  \\    
$m \alpha^5$      &      -1.06(4)  &      -1.0106(10)&      -0.46(7)  &      -1.73(5)  &      -2.46(5)  &      -0.61(5)   \\             
$m \alpha^6$      &      -0.048(10)&      -0.045(9)  &      -0.021(5) &      -0.08(2)  &      -0.11(3)  &      -0.027(7)\\[1ex]             
Total             & 21\,980.18(5)  & 42\,565.441(11) & 52\,080.99(7)  & 56\,882.55(6)  & 59\,696.07(6)  & 62\,053.79(6)  \\[1ex]
Theory (ECG)      & 21\,979.4(11)^e   \\
Theory (FCPC)$^f$ & 21\,980.85     & 42\,568.80      & 52\,081.09     & 56\,890.9      & 59\,699.8      & 62\,055.25 \\
Theory (MCHF)$^h$ & 22\,099.30     & 42\,710.97      & 52\,080.09     & 56\,945.89     & 59\,793.35     & 62\,165.91 \\
Experiment        & 21\,980.16(8)^{a,g} &  42\,565.450\,2(10)^b & 52\,080.94(6)^a & 56\,882.547\,4(21)^c & 59\,696.07(5)^{a,g}       & 62\,053.74(6)^{a,g}\\
Total $-$ Experiment$^d$& 0.02(5) &	-0.009(11) & 0.05(7) & 0.00(6) & 0.04(6) & 0.05(6) \\[2ex]
Total (ionization)   &    53\,212.51(5)   &    32\,627.265(11)       &  23\,111.73(5)     &     18\,310.13(5)         &       15\,496.62(6)    &   13\,138.90(5)    \\[1ex]
Theory(FCPC)$^f$            &    53\,211.22      &    32\,623.27      &  23\,110.98        &                      &                        &   13\,136.8         \\
\end{tabular}
\end{ruledtabular}
\begin{flushleft}
$^a$ Johansson \cite{Johansson:62}, Kramida \etal \cite{Kramida:20};
$^b$ Cook \etal \cite{Cook:18};
$^c$ Cook \etal \cite{Cook:20};
$^d$ Theoretical uncertainty assumed.;
$^e$ Kedziorski \etal~\cite{Kedziorski:20}, averaged over $J$;
$^f$ Chung and Zhu \cite{Chung:93,Zhu:95};
$^g$ Centroid uncertainty taken as the maximum error from the individual $J$ lines.;
$^h$ Fischer and Tachiev \cite{Fischer:04}.
\end{flushleft}
\end{table*}

\subsection{Comparison with experimental and other theoretical results}

Table~\ref{Tcentr} presents our recommended total energies of centroids and their components 
of Eq.~(\ref{eq:expansion}). The energy of the ground $2s^2\,^1\!S$ level was taken as a reference for 
the wavenumber scale. The extrapolated values and uncertainties of the nonrelativistic 
contribution (see Table~\ref{Tconv}) and of all the corrections
were estimated from their convergence with increasing size of basis sets.
As can be inferred from the table, except for the $2s2p\,^1\!P$ state, that the accuracy 
of the total energy is limited mainly by the uncertainty of the leading QED correction. 
This correction is dominated by the Bethe logarithm term, which is estimated using 
$\ln{k_0}=5.752(3)$ \cite{Puchalski:13a}.

The total energy of a centroid for all the six states considered here is consistent 
within the theoretical uncertainty with the experimental one 
\cite{Johansson:62,Kramida:20,Cook:18,Cook:20}---see 'Total $-$ Experiment' entries 
in Table~\ref{Tcentr}. The agreement is on the level of $0.01\,\icm$. 
In particular, our predictions agree with order-of-magnitude more accurate
measurements reported by Williams group \cite{Cook:18,Cook:20}.
For the $2p^2\,^3\!P^e_J$ term, there is a small difference between the original triplet 
($J=0,1,2$) wavenumbers by Johansson \cite{Johansson:62} and those reported on the NIST web page
\cite{Kramida:20}, which affects the centroid values. The experimental value placed 
in Table~\ref{Tcentr} refers to original measurements and we note that value from 
the NIST compilation is smaller by $0.09\,\icm$.

There are also scarce theoretical data in the literature concerning selected excited states of $^9$Be.
Chung and Zhu \cite{Chung:93,Zhu:95} evaluated the energy using the FCPC 
method and included the relativistic and QED correction but without the uncertainty estimation, 
see Table~\ref{Tcentr}. 
Their centroid energies differ from ours and from the experimental ones by 0.1 to 3.4 $\icm$.
Much newer results obtained by Fischer and Tachiev using a multiconfiguration Hartree-Fock (MCHF)
\cite{Fischer:04} method differ from ours by as much as $63-146\,\icm$ with the exception of the
$^3\!S$ level (\hbox{$\sim\!1\,\icm$}).
The centroid energy of the $2s2p\,^3\!P$ state obtained from Kedziorski's calculations~\cite{Kedziorski:20},
being less accurate, agrees within the uncertainty with our result. Surprisingly,
their relativistic correction $12.40(7)$ cm$^{-1}$ is in significant disagreement with our value 
of $13.189(3)$ cm$^{-1}$. The difference between these two values corresponds to the difference between
their centroid energy and the experimental one, and thus raises doubts about their uncertainty estimation. 

\begin{figure}[!ht]
\includegraphics[scale=0.45]{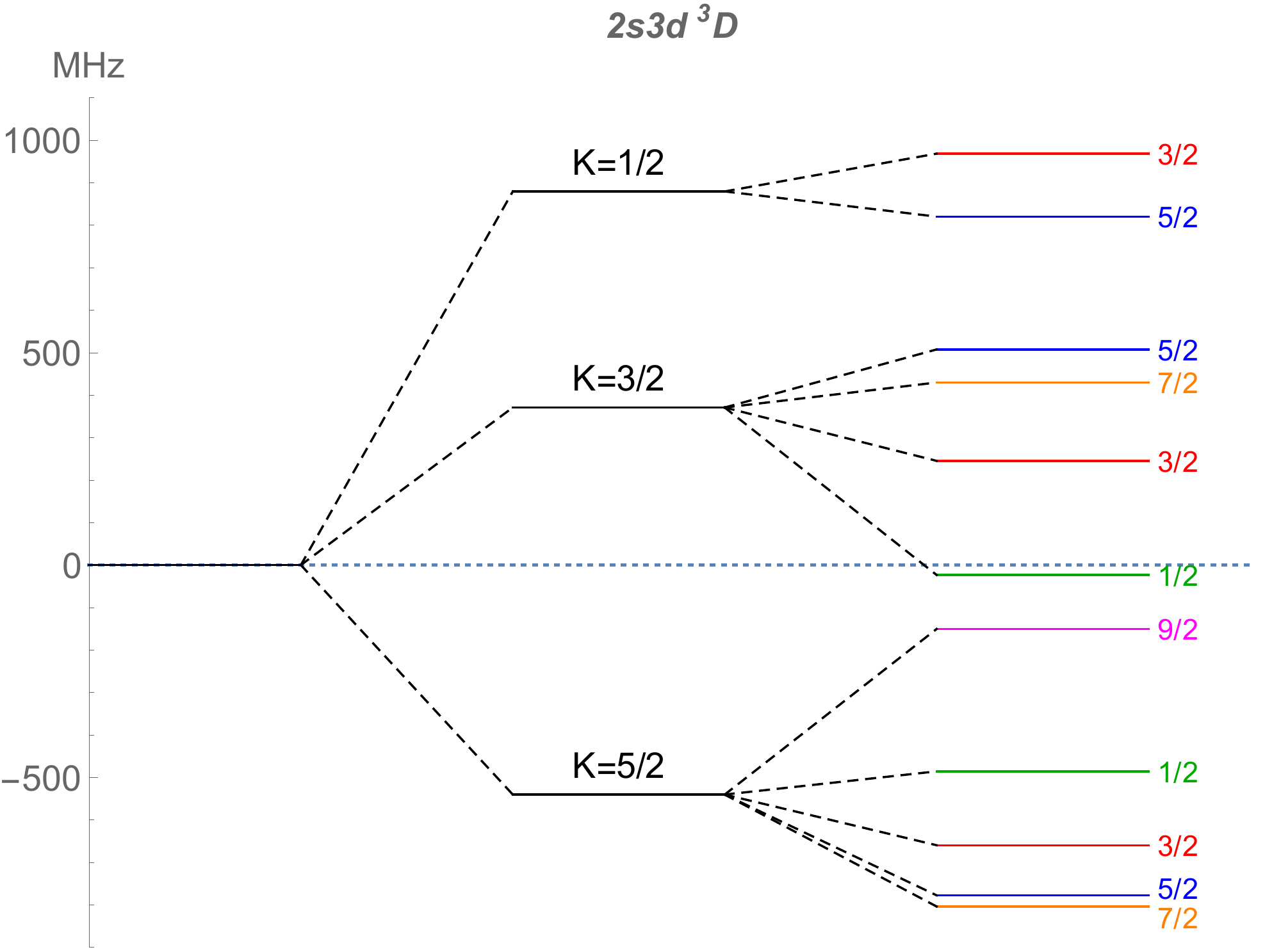}
\caption{The fine and hyperfine splitting for the $2s3d\,^3\!D$ state. The angular momentum number $K$, defined as $\vec K = \vec I+\vec S$,
is an approximate quantum number.}
\label{F3D}
\end{figure}

Fine-structure splittings obtained theoretically and experimentally for both $^3\!P$ terms
agree well with each other. Separate comment is required concerning the $2s3d\,^3\!D$ term.
Its fine-structure has not been revealed in Johansson's experiments \cite{Johansson:62}.
Even then, it has been given on the NIST page \cite{Kramida:20}. The fine splitting has also
been predicted theoretically by Chung and Zhu \cite{Chung:93} and by Fischer and Tachiev
\cite{Fischer:04}. However, in view of the clear domination of the $IS$ over $LS$ coupling 
(see discussion in Sec.~\ref{sec:fshfs}), we claim that for this term the notion 
of the fine-structure should be either totally abandoned or at least reinterpreted in terms 
of the $\vec{K}=\vec{I}+\vec{S}$ angular momentum (see Fig.~\ref{F3D}).

\begin{table*}[!hbt]
\renewcommand{\arraystretch}{1.0}
\caption{The nonrelativistic energy and theoretical fine and hyperfine structure
parameters for the $2s2p\,^3\!P$ state of $^9$Be (in MHz). Nuclear mass 
$m_\mathrm{N}= 9.012\,183\,07(8)$ u \cite{Wang:17}, magnetic moment $\mu/\mu_{\rm N} = -1.177\,432(3)$
\cite{Stone:15}, and quadrupole moment $Q_\mathrm{N}=0.05350(14)$ barns \cite{Puchalski:21} were used 
for the $^9$Be nucleus. The $a_1$ coefficient is to be multiplied by $1+\epsilon$ according to Eq. (\ref{16}).
Shown uncertainties are of the numerical origin, while implicit relative uncertainties due to unknown relativistic corrections are about $\alpha^2$} 
\label{Tabcparms}
\begin{ruledtabular}
\begin{tabular}{rw{6.4}w{5.5}w{4.7}w{4.8}w{3.7}w{2.9}}
State & \cent{c_1}  & \cent{c_2}  & \cent{a_1}  & \cent{a_2}  &  \cent{a_3}  & 
\cent{b/Q_\mathrm{N}\,(\mathrm{MHz/barn})} \\
\hline
$2s3s\,^3\!S$   &                &                & -332.283\,4(3) &                     &                &               \\    
$2s2p\,^1\!P$   &                &                &                & -13.888\,28(8)      &                & 15.801\,75(7) \\
$2s2p\,^3\!P$   & 32\,987.6(2)   &  5\,399.15(2)  & -231.128\,4(6) & -22.699\,24(5)      &  14.788\,68(5) & 27.148\,87(3) \\
$2p^2\,^3\!P^e$ & 32\,321.08(9)  & -3\,973.643(4) &   34.732\,27(8)& -23.263\,96(15)     & -14.757\,164(4)&-27.424\,71(4) \\
$2p2p\,^1\!D$   &                &                &                & -13.963\,93(3)      &                & 32.575\,5(3)  \\
$2s3d\,^3\!D$   &      90.400(9) &    124.654\,6(2) & -303.169\,8(6) &  -0.404\,656\,7(15) &  0.078\,38(7)&  0.838\,9(6)  \\
\end{tabular}
\end{ruledtabular}
\end{table*}

\begin{table}[!hbt]
\renewcommand{\arraystretch}{1.1}
\caption{Fine/hyperfine levels (in MHz) of low lying excited states of $^9$Be atom. 
$\vec J= \vec L+\vec S$, $\vec K = \vec I+\vec S$, $\vec{F}=\vec{L}+\vec{S}+\vec{I}$.
The numerical uncertainty is negligible, while the implicit relative uncertainty due to
unknown  higher order relativistic corrections is about $\alpha^2$.}
\label{Thfslevels}
\begin{ruledtabular}
\begin{tabular}{lw{6.3}w{6.3}lw{4.4}}
\cent{\nu_J(F)} & \cent{2s2p\,^3\!P} & \cent{2p^2\,^3\!P^e} & \cent{\nu_K(F)}  & \cent{2s3d\,^3\!D}\\
\hline
$\nu_0(3/2)$    &  -56\,982.      & -71\,265.          & $\nu_{1/2}(3/2)$ &  968.54 \\	 
$\nu_1(1/2)$    &  -37\,140.      & -29\,054.          & $\nu_{1/2}(5/2)$ &  820.31 \\
$\nu_1(3/2)$    &  -37\,343.      & -29\,028.          & $\nu_{3/2}(1/2)$ &  -23.38 \\     
$\nu_1(5/2)$    &  -37\,697.      & -28\,983.          & $\nu_{3/2}(3/2)$ &  245.25 \\     
$\nu_2(1/2)$    &   34\,449.      &  31\,643.          & $\nu_{3/2}(5/2)$ &  507.55 \\	  
$\nu_2(3/2)$    &   34\,262.      &  31\,649.          & $\nu_{3/2}(7/2)$ &  429.71 \\
$\nu_2(5/2)$    &   33\,950.      &  31\,658.          & $\nu_{5/2}(1/2)$ & -485.73 \\     
$\nu_2(7/2)$    &   33\,514.      &  31\,668.          & $\nu_{5/2}(3/2)$ & -659.71 \\   
                &                    &                  & $\nu_{5/2}(5/2)$ & -777.63 \\	   
                &                    &                  & $\nu_{5/2}(7/2)$ & -803.93 \\
                &                    &                  & $\nu_{5/2}(9/2)$ & -150.57 \\[2ex]
\cent{\nu_J(F)} & \cent{\qquad 2s2p\,^1\!P_1} & \cent{\quad 2s3s\,^3\!S_1} & \cent{\nu_J(F)}  & \cent{2p^2\,^1\!D_2}\\
\hline
$\nu_1(1/2)$    &   35.777          &  831.25             & $\nu_2(1/2)$     &  64.363 \\	 
$\nu_1(3/2)$    &   13.043          &  332.50             & $\nu_2(3/2)$     &  41.892 \\
$\nu_1(5/2)$    &  -20.621          & -498.75             & $\nu_2(5/2)$     &   5.893 \\
                &                   &                     & $\nu_2(7/2)$     & -41.456 \\
\end{tabular}                                                
\end{ruledtabular}
\end{table}
                          
%
                 
\section{Conclusions}

We have performed the most accurate calculations of centroid energies and fine/hyperfine structure
parameters of low lying $^{1,3}\!P_J$, $^{3}\!S_1$, $^{3}\!P^e_J$, and $^{1,3}\!D_J$ excited states 
of the $^9$Be atom. The obtained results, apart from being in agreement with available experimental values, 
allow the accuracy of standard atomic structure calculations to be assessed.
For a long time, the FCPC method by Chung and Zhu \cite{Chung:93} was considered the most 
accurate one regarding the centroid energies which included the finite nuclear mass,
relativistic, and QED corrections. 
However, significant differences between the FCPC results and our calculations (and experiments), 
reaching several reciprocal centimeters  (up to $8\,\icm$ for the $^1\!D$ state), show the importance of
the use of high-quality wave functions in accurate studies of the atomic structure. 
It is noteworthy that these differences are often greater than the entire QED correction. 
What is more, some of the previous calculations of the combined fine and hyperfine structure 
were performed incorrectly. The reason was that the $LS$ coupling was assumed to be dominant 
over all other couplings. In other words, we demonstrated that the standard approach 
in terms of $A_J$ and $B_J$ hyperfine parameters does not work for some $^9$Be excited states, 
and that a subtle analysis is necessary to properly identify the origin of energy level splitting.

Apart from presenting accurate results for the fine and hyperfine structure, we have introduced 
an approach allowing us to handle the combined fine and hyperfine structure of an arbitrary 
atomic system in terms of an effective Hamiltonian. This Hamiltonian is to be diagonalized 
for particular values of the fine/hyperfine coupling parameters. This approach is particularly 
suitable in cases where the hyperfine mixing becomes significant. 
Moreover, we expressed the fine and hyperfine parameters in terms of reduced matrix elements 
with Cartesian angular factors. These factors can conveniently be combined with a general 
correlated basis function and applied to an arbitrary atomic term.

The current capabilities of theoretical methods are limited by the accuracy of both 
$m\alpha^5$ and $m\alpha^6$ QED components. As shown by the instance of the $2s2p\,^1\!P$ state,
more accurate calculations of the former component are feasible, but,
significantly more effort will be needed to evaluate accurately the latter component. 
Its complete evaluation will require construction of wave functions strictly obeying 
the cusp condition as it has already been demonstrated for two-electron systems 
\cite{Puchalski:17,Puchalski:19}.

\section*{Acknowledgments}
Fruitful exchange of information with Alexander Kramida is acknowledged.
This research was supported by National Science Center (Poland) Grants No. 2014/15/B/ST4/05022, 2019/34/E/ST4/00451 
and 2017/27/B/ST2/02459, as well as by a computing grant from Pozna\'n Supercomputing and Networking
Center and by PL-Grid Infrastructure.

\appendix

\section{Regularization}
\label{App:regularization}

This appendix briefly describes the regularization technique of singular operators used in this work. 
We can assume that a operator $Q$ to be regularized depends only on a two-particle 
coordinate $\vec r_{aX}$, i.e. electron-nucleus $Q=Q(\vec r_a)\equiv Q(\vec r_{aN})$, $X=N$
or electron-electron $ Q= Q(\vec r_{ab})$, $X=b$.

For any operator $Q$, one finds a corresponding  operator $\tilde{Q}$, such that 
\begin{align}
 \bigg(\frac{1}{m} + \frac{1}{m_X}\bigg)\,Q =&\ \frac{1}{m_\mathrm{N}} \big[\vec p_\mathrm{N}\,,\big[\vec p_\mathrm{N}\,,\tilde Q\big]\big] + \frac{1}{m}\,\sum_{c} \big[\vec p_c\,,\big[\vec p_c\,,\, \tilde{Q}\big]\big] 
\nonumber \\ =&\ 
[Q]_r - 2\,\big\{E_0-H^{(2)},\tilde{Q} \bigr\}\,,
\end{align}
where the curly bracket denotes an anticommutator and
\begin{equation}
[Q ]_r = 4\,(E_0 - V)\,\tilde{Q} - \sum_c \vec p_c\,\tilde{Q}\,\vec p_c - \frac{1}{m_\mathrm{N}}\,\vec p_\mathrm{N}\,\tilde{Q}\,\vec p_\mathrm{N} \,.
\end{equation}
Using the above notation, an expectation value of a one-electron operator $Q=Q(\vec r_a)$
can be represented in the regularized form as 
\begin{equation}
 \langle\Psi|Q|\Psi\rangle =  \bigg(\frac{1}{m} + \frac{1}{m_X}\bigg)^{-1}  \langle\Psi|[Q]_r|\Psi\rangle
\end{equation}
There are two one-electron operators, i.e. $4\,\pi\,\delta^3(r_{a})$ and 
$r_a^{-5}\,(r_a^i\,r_a^j - 1/3\,\delta^{ij}\,r_a^2)$, and one two-electron operator, i.e.  
$4\,\pi\,\delta^3(r_{ab})$, to be regularized. The corresponding $\tilde{Q}$ operators are 
of the following form:  $r_a^{-1}$, $1/6\,r_a^{-3}\,(r_a^i\,r_a^j - \delta^{ij}\,r_a^2/3)$ 
and $r_{ab}^{-1}$, respectively. \\

Another regularization scheme is needed for $\sum_{a} \vec p_a^{\,4}$. In this case
\begin{align}
\sum_{a} \vec p_a^{\,4} &= [\sum_{a} \vec p_a^{\,4}]_r - 4\,\big\{E_0-H^{(2)},V \bigr\}\,,
\intertext{where}
\sep[\sum_{a} \vec p_a^{\,4}]_r &= 4 \,(E_0-V)^2 - 2\, \sum_{a<b} \vec p_a^{2} \, \vec p_b^{2} \\\nonumber
&\quad - \frac{4}{m_\mathrm{N}}\,(E_0-V)\,\vec p_\mathrm{N}^{2} + \frac{1}{m_\mathrm{N}^2}\,\vec p_\mathrm{N}^{\,4}
\end{align}
Using this definition the following identity can be written
\begin{eqnarray}
\langle\Psi|\sum_{a} \vec p_a^{\,4}|\Psi\rangle &=& \langle\Psi| [\sum_{a} \vec p_a^{\,4}]_r |\Psi \rangle\,.
\end{eqnarray}

Finally, we make use of the following regularization scheme for
the Araki-Sucher term (in the infinite mass limit)~\cite{Pachucki:04} 
\begin{align}
\bigg \langle \Psi \bigg| P\left(\frac{1}{r_{ab}^3}\right) \bigg| \Psi \bigg\rangle &=
\sum_c \bigg \langle \Psi \bigg| \vec{p}_c\,\frac{\ln r_{ab}}{r_{ab}}\, \vec{p}_c \bigg| \Psi \bigg \rangle \\
& \hspace{-1cm} 
+\bigg \langle  \Psi \bigg| 4 \pi\, (1 + \gamma)\, \delta(r_{ab}) + 2\, (E_0 - V) \,\frac{\ln r_{ab}}{r_{ab}} \bigg| \Psi \bigg \rangle\,. \nonumber
\label{reg_rabm3}
\end{align}

\section{Table of u- coefficients} 
\label{App:u-coeff}
\begin{table*}[!hbt]
\renewcommand{\arraystretch}{1.0}
\caption{Matrix elements reduction coefficients, see Sec.~\ref{Sec:Rscalar}.}
\label{u-coefficients}
\begin{ruledtabular}
\begin{tabular}{lrrrrrrrrrrrr}
$\mathcal{P}_l$  & $2\,u_{l}$(singlet) & $2\,u_{l}$ & $2\,u^{1}_{l}$ & $2\,u^{2}_{l}$ & $2\,u^{3}_{l}$& $2\,u^{4}_{l}$ & 
$2\,u^{12}_{l}$ & $2\,u^{13}_{l}$ & $2\,u^{14}_{l}$& $2\,u^{23}_{l}$ & $2\,u^{24}_{l}$& $2\,u^{34}_{l}$
\\
\hline 
 \\
 $1234$   &  2   &   2   &  0   & 0   &  2  &  2   &  0   &  0   &  0   &  0   &  0   &  -2 \\
 $1243$   &  2   &   -2  &  0   & 0   &  -2 &  -2  &  0   &  0   &  0   &  0   &  0   &  2   \\
 $1324$   &  -1  &   -1  &  1   & -1  &  -1 &  -1  &  0   &  0   &  -1  &  0   &  1   &  1  \\
 $1342$   &  -1  &   1   &  -1  & 1   &  1  &  1   &  0   &  0   &  1   &  0   &  -1  &  -1  \\
 $1423$   &  -1  &   1   &  -1  & 1   &  1  &  1   &  0   &  1   &  0   &  -1  &  0   &  -1  \\
 $1432$   &  -1  &   -1  &  1   & -1  &  -1 &  -1  &  0   &  -1  &  0   &  1   &  0   &  1  \\
 $2134$   &  2   &   2   &  0   & 0   &  2  &  2   &  0   &  0   &  0   &  0   &  0   &  -2 \\
 $2143$   &  2   &   -2  &  0   & 0   &  -2 &  -2  &  0   &  0   &  0   &  0   &  0   &  2  \\
 $2314$   &  -1  &   -1  &  -1  & 1   &  -1 &  -1  &  0   &  0   &  1   &  0   &  -1  &  1   \\
 $2341$   &  -1  &   1   &  1   & -1  &  1  &  1   &  0   &  0   &  -1  &  0   &  1   &  -1 \\
 $2413$   &  -1  &   1   &  1   & -1  &  1  &  1   &  0   &  -1  &  0   &  1   &  0   &  -1 \\
 $2431$   &  -1  &   -1  &  -1  & 1   &  -1 &  -1  &  0   &  1   &  0   &  -1  &  0   &  1   \\
 $3124$   &  -1  &   -1  &  1   & -1  &  -1 &  -1  &  0   &  0   &  -1  &  0   &  1   &  1  \\
 $3142$   &  -1  &   1   &  -1  & 1   &  1  &  1   &  0   &  0   &  1   &  0   &  -1  &  -1  \\
 $3214$   &  -1  &   -1  &  -1  & 1   &  -1 &  -1  &  0   &  0   &  1   &  0   &  -1  &  1   \\
 $3241$   &  -1  &   1   &  1   & -1  &  1  &  1   &  0   &  0   &  -1  &  0   &  1   &  -1 \\
 $3412$   &  2   &   0   &  0   & 0   &  0  &  0   &  0   &  1   &  -1  &  -1  &  1   &  0  \\
 $3421$   &  2   &   0   &  0   & 0   &  0  &  0   &  0   &  -1  &  1   &  1   &  -1  &  0  \\
 $4123$   &  -1  &   1   &  -1  & 1   &  1  &  1   &  0   &  1   &  0   &  -1  &  0   &  -1  \\
 $4132$   &  -1  &   -1  &  1   & -1  &  -1 &  -1  &  0   &  -1  &  0   &  1   &  0   &  1  \\
 $4213$   &  -1  &   1   &  1   & -1  &  1  &  1   &  0   &  -1  &  0   &  1   &  0   &  -1 \\
 $4231$   &  -1  &   -1  &  -1  & 1   &  -1 &  -1  &  0   &  1   &  0   &  -1  &  0   &  1   \\
 $4312$   &  2   &   0   &  0   & 0   &  0  &  0   &  0   &  1   &  -1  &  -1  &  1   &  0  \\
 $4321$   &  2   &   0   &  0   & 0   &  0  &  0   &  0   &  -1  &  1   &  1   &  -1  &  0  \\
  \end{tabular}
\end{ruledtabular}
\end{table*}


\end{document}